\documentclass[12pt,preprint]{aastex}
\newcommand{\etal}{et~al.}
\newcommand{\CIVdblt}{{\rm C}\kern 0.1em{\sc iv}~$\lambda\lambda 1548, 1550$}
\newcommand{\MgIIdblt}{{\rm Mg}\kern 0.1em{\sc ii}~$\lambda\lambda 2796, 2803$}
\newcommand{\SiIVdblt}{{\rm Si}\kern 0.1em{\sc iv}~$\lambda\lambda 1393, 1402$} 
\newcommand{\NVdblt}{\hbox{{\rm N}\kern 0.1em{\sc v}~$\lambda\lambda 1239,1243$}}
\newcommand{\OVIdblt}{{\rm O}\kern 0.1em{\sc vi}~$\lambda\lambda 1032, 1038$} 
\newcommand{\CII}{\hbox{{\rm C}\kern 0.1em{\sc ii}}}
\newcommand{\CIII}{\hbox{{\rm C}\kern 0.1em{\sc iii}}}
\newcommand{\CIV}{\hbox{{\rm C}\kern 0.1em{\sc iv}}}
\newcommand{\HI}{\hbox{{\rm H}\kern 0.1em{\sc i}}}
\newcommand{\NaI}{\hbox{{\rm Na}\kern 0.1em{\sc i}}}
\newcommand{\HII}{\hbox{{\rm H}\kern 0.1em{\sc ii}}}
\newcommand{\HeI}{\hbox{{\rm He}\kern 0.1em{\sc i}}}
\newcommand{\HeII}{\hbox{{\rm He}\kern 0.1em{\sc ii}}}
\newcommand{\Lya}{\hbox{{\rm Ly}\kern 0.1em$\alpha$}}
\newcommand{\Lyb}{\hbox{{\rm Ly}\kern 0.1em$\beta$}}
\newcommand{\Lyg}{\hbox{{\rm Ly}\kern 0.1em$\gamma$}}
\newcommand{\Lyd}{\hbox{{\rm Ly}\kern 0.1em$\delta$}}
\newcommand{\Lye}{\hbox{{\rm Ly}\kern 0.1em$8$}}
\newcommand{\FeII}{\hbox{{\rm Fe}\kern 0.1em{\sc ii}}}
\newcommand{\MgI}{\hbox{{\rm Mg}\kern 0.1em{\sc i}}}
\newcommand{\MgII}{\hbox{{\rm Mg}\kern 0.1em{\sc ii}}}
\newcommand{\OVI}{\hbox{{\rm O}\kern 0.1em{\sc vi}}}
\newcommand{\OVII}{\hbox{{\rm O}\kern 0.1em{\sc vii}}}
\newcommand{\OVIII}{\hbox{{\rm O}\kern 0.1em{\sc viii}}}
\newcommand{\NV}{\hbox{{\rm N}\kern 0.1em{\sc v}}}
\newcommand{\SiII}{\hbox{{\rm Si}\kern 0.1em{\sc ii}}}
\newcommand{\SiIII}{\hbox{{\rm Si}\kern 0.1em{\sc iii}}}
\newcommand{\SiIV}{\hbox{{\rm Si}\kern 0.1em{\sc iv}}}
\newcommand{\kms}{\hbox{km~s$^{-1}$}}
\newcommand{\cmsq}{\hbox{cm$^{-2}$}}

\tighten
\begin{document}
 
 


\title{The Absorption Signatures of Dwarf Galaxies: The $\lowercase{z}=1.04$ Multi--cloud Weak {\MgII} Absorber Toward PG~1634+706: 
\altaffilmark{1,2}}

\author{Stephanie~G.~Zonak\altaffilmark{3}, Jane~C.~Charlton\altaffilmark{4},
Jie~Ding\altaffilmark{4}, and Christopher~W.~Churchill\altaffilmark{5,6}}
\affil{Department of Astronomy and Astrophysics \\ The Pennsylvania
State University \\ University Park, PA 16802 \\ {\it szonak,
charlton, ding, cwc@astro.psu.edu}}

\altaffiltext{1}{Based  in  part   on  observations  obtained  at  the
W.~M. Keck Observatory, which  is operated as a scientific partnership
among Caltech, the University of California, and NASA. The Observatory
was made possible by the  generous financial support of the W.~M. Keck
Foundation.}
\altaffiltext{2}{Based  in  part  on  observations obtained  with  the
NASA/ESA {\it Hubble Space Telescope},  which is operated by the STScI
for the  Association of Universities for Research  in Astronomy, Inc.,
under NASA contract NAS5--26555.}
\altaffiltext{3}{Department of Astronomy, Building 225, College Park, MD 20742}
\altaffiltext{4}{Department of Astronomy and Astrophysics, The Pennsylvania
State University, University Park, PA 16802, {\it
charlton, ding@astro.psu.edu}}
\altaffiltext{5}{Department of Astronomy New Mexico State University
  1320 Frenger Mall, Las Cruces, New Mexico 88003-8001
{\it cwc@nmsu.edu}}
\altaffiltext{6}{Visiting Astronomer at the W.~M. Keck Observatory}

\begin{abstract}
We analyze high resolution spectra of a multi--cloud weak [defined as
$W_r({\MgII}) < 0.3$~{\AA}] absorbing system along the line of sight to
PG~$1634+706$.  This system gives rise to a partial Lyman limit break
and absorption in {\MgII}, {\SiII}, {\CII}, {\SiIII}, {\SiIV}, {\CIV},
and {\OVI}.  The lower ionization transitions arise in two kinematic
subsystems with a separation of $\simeq 150$~{\kms}.  Each subsystem
is resolved into several narrow components, having Doppler widths of
$3$--$10$~{\kms}.  For both subsystems, the {\OVI} absorption arises
in a separate higher ionization phase, in regions dominated by bulk
motions in the range of $30$--$40$~{\kms}.  The two {\OVI} absorption
profiles are kinematically offset by $\simeq 50$~{\kms} with respect
to each of the two lower ionization subsystem.  In the stronger
subsystem, the {\SiIII} absorption is strong with a distinctive,
smooth profile shape and may partially arise in shock heated gas.
Moreover, the kinematic substructure of {\SiIV} traces that of the
lower ionization {\MgII}, but may be offset by $\simeq 3$~{\kms}.

Based upon photoionization models, constrained by the partial Lyman
limit break, we infer a low metallicity of $\sim 0.03$ solar for the
low ionization gas in both subsystems.  The broader {\OVI} phases have
a somewhat higher metallicity, and they are consistent with
photoionization; the profiles are not broad enough to imply production
of {\OVI} through collisional ionization.  Various models, including
outer disks, dwarf galaxies, and superwinds, are discussed to account
for the phase structure, metallicity, and kinematics of this
absorption system.  We favor an interpretation in which the two
subsystems are produced by condensed clouds far out in the opposite
extremes of a multi--layer dwarf galaxy superwind.

\end{abstract}
%
%


\section{Introduction}
\label{sec:intro}

Quasar absorption line systems, as traced by their {\MgII}, provide a
unique way to study our universe.  Current classifications schemes for
systems with detected {\MgII} file them into three main categories:
those that are characterized by their strong {\MgII} absorption, those
that reveal weak, narrow, single--cloud {\MgII} profiles, and those
that exhibit weak, multiple cloud {\MgII} absorption.  In this paper
we analyze in detail the absorption profiles from a particular
multiple cloud weak {\MgII} absorber at $z=1.04$ along the line of
sight toward PG~$1634+706$.  The longer--term goal of a collection of
similar studies will be to understand the relationships between the
different classes of absorption systems, and the connections with the
different types of galaxies and structures at various redshifts.

Although there are no distinct divisions between the three categories
of {\MgII} absorption systems, by their properties they appear to
be related to three different types of gaseous structures at
$z\sim1$.  Strong {\MgII} absorbers [those with $W_r(2796)>0.3$~{\AA}]
show Lyman limit breaks and contain multiple clouds spread over tens
to hundreds of kilometers per second \citep{archiveI}.  A large
majority of these absorbers are known to be associated with luminous
galaxies ($>0.05 L^*$, where $L^*$ is the Schechter luminosity),
within an impact parameter of $40h^{-1}$~kpc of the quasar
\citep{bb91,bergeron92,lebrun93,sdp94,steidel95,3c336}.
All of the strong {\MgII}
absorbers ($dN/dz = 0.91 \pm 0.1$ at $<z>=0.9$ \citep{ss92}) can be
accounted for by regions of this size around the known population of
luminous galaxies.  The strong {\MgII} absorber spectral profiles, in
chemical transitions of low and high ionization states, generally
require multiple phases of gas, i.e. regions of differing densities
that are spatially distinct (e.g. \citep{ding1634,ding1206}).  The
kinematics of the low ionization gas is generally consistent with what
one would expect from the combined disks and halos of galaxies of a
variety of morphological types \citep{kinmod,steidel02}.

In contrast to the strong {\MgII} absorbers, the single--component
systems with weaker {\MgII} (rest frame equivalent widths $W(2796)
<0.3$~{\AA}) are typically not known to be directly associated with
luminous galaxies (i.e. they are not within $50$--$100$~kpc of $>0.05
L^*$ galaxies) \citep{weak1}.  These single--component weak {\MgII}
systems have a significant absorption cross-section, with $dN/dz =
1.10 \pm0.06$ for $0.02< W< 0.3$~{\AA} at $0.4<z<1.0$ \citep{weak1},
similar to the value for all strong {\MgII} absorbers.  The
distribution of the number of Voigt profile components used to fit
{\MgII} absorption systems is roughly Gaussian with a median of seven
components.  However, there is a strong excess of single cloud
components relative to the rest of the distribution (see Fig.~2 in
\citet{weak2} and Table 7 of \citet{cv01}).
This implies that the single--component weak {\MgII} absorbers are
primarily produced by a different class of object than the strong
{\MgII} absorbers.  Consistent with this interpretation, most of the
single--component weak {\MgII} absorbers do not produce Lyman limit
breaks \citep{archiveI}.  Their {\MgII} profiles are generally quite
narrow, with Doppler parameters of $b\sim 2$ -- $5$~{\kms}.  Their
{\CIV} profiles are broader and require a separate, higher ionization,
phase of gas.  Their low ionization phases are inferred to have
metallicities greater than a tenth solar and to have sizes less than
$\sim 10$--$100$~pc \citep{weak2,weak1634}.  The general properties of
the single--cloud weak {\MgII} absorbers suggest an origin in some
type of faded early extragalactic star cluster or in metal--rich fragments
in cold dark matter mini--halos ''failed galaxies'' \citep{weak2}.

The third category of {\MgII} systems is characterized by multiple,
weak components spread over tens of kilometers per second. About one
third of the weak {\MgII} absorbers are in this category \citep{weak1}. By
definition, these are ``weak {\MgII} absorbers'', since the total rest
frame equivalent width of {\MgII}~2796 is $<0.3$~{\AA}.  Unlike the
the strong {\MgII} absorbers, little is directly known about the
environments of the multiple--component weak {\MgII} systems.  However,
the multiple--component weak category appears consistent with an
extension of the distribution of the numbers of components fit to
strong {\MgII} absorbers (see Fig.~2 in \citet{weak2}).  The weakness
and kinematics of the multiple--component weak {\MgII} profiles might
suggest a line of sight through the outer regions of a galaxy where
the gas could be more diffuse and less dense.  It is also possible
that a dwarf galaxy, or even a low--metallicity giant galaxy would
give rise to such weak {\MgII} absorption.

In this paper, we study a multiple--component, weak {\MgII} [rest
frame equivalent width, $W_r(2796) =0.097
\pm 0.008$~{\AA}] absorber at $z\sim1$.  
This absorber was modeled previously by \citet{anticip}, using
low--resolution FOS/HST data along with the high--resolution {\MgII}
profiles from HIRES/Keck.  They determined that the {\MgII} clouds
have a relatively low metallicity of $0.03$ solar, based on a comparison to a
partial Lyman limit break.  The {\CIV}, detected in the FOS spectrum,
was clearly offset from the low ionization clouds.  This required an
additional high ionization cloud producing absorption $\sim 200$~{\kms} to the
blue of the detected {\MgII} clouds.  \citet{anticip} predicted that
this cloud should produce observable {\CII}, {\SiIII}, and {\SiIV} in
the high resolution STIS spectrum.  However, with just the low--resolution
UV spectrum, they were unable to determine if the {\CIV} detected at
the same velocity as {\MgII} could arise in the same phase of gas with
the {\MgII}, or if a separate phase would be needed.  Also, the ionization
conditions and metallicities of the high ionization clouds could not
be well--constrained.

STIS/HST spectra of PG~$1634+706$ are now available, with high
resolution coverage of {\Lya} and the Lyman series, {\SiII}, {\SiIII},
{\CII}, {\SiIV}, and {\OVI}.  With these new data, we aim to test the
models of \citet{anticip}, and to reach more detailed conclusions
about the phase structure, metallicity, ionization conditions and
kinematic properties of the $z=1.04$ system toward PG~$1634+706$.

In \S~\ref{sec:data} we describe the spectra of quasar PG~$1634+706$ used
in our analysis, obtained with STIS/HST, FOS/HST and HIRES/Keck.  We
then present the data for the $z=1.04$ system along this line of sight
in \S~\ref{sec:system} and make a qualitative comparison of the
individual chemical transitions.  Our analysis method, considering
photoionization and collisional ionization models, is outlined in
\S~\ref{sec:method}.  In \S~\ref{sec:results} we present 
results from modeling of this system and we summarize the conclusions
in \S~\ref{sec:summary}. 
In \S~\ref{sec:discussion},
we discuss possible scenarios for the origin of the $z=1.04$ system
and the general implications for the nature of multiple--component,
weak {\MgII} absorbers.


\section{Data}
\label{sec:data}
Spectra of the $z= 1.36$ quasar PG~$1634+706$, obtained with the three
instruments HIRES/Keck, STIS/HST, and FOS/HST, were used to study the
various transitions from an intervening absorber at $z=1.0414$.  Other
{\MgII} systems along this line of sight are published
elsewhere. There are three weak, single--cloud {\MgII} absorbers at
$z=0.6540$, $z=0.8181$ and $z=0.9056$ \citep{weak1,weak2,weak1634} and
a strong {\MgII} absorber at $z=0.9902$ \citep{cv01,ding1634}.

Although a WFPC2/HST image of the quasar field exists, it is not deep,
so it provides only a limited constraint on the luminosity of the host
galaxy of this $z=1.04$ absorber.  After subtraction of the point
spread function of the $V=14.9$ quasar from the image, \citet{farrah}
did not detect any galaxies within the $\sim 36\arcsec$ Planetary
Camera field of view to a magnitude of $m_I = 22.5$ (at the $3\sigma$
level).  This implies that there are no galaxies brighter than $\sim
0.5L^*$ with impact parameter less than $\sim 500$~kpc at the redshift
of our absorber (\citet{farrah} used $h = 0.65$).  Due to
uncertainties in the PSF the limit is not as restrictive within $1 -
2$ kpc.  Since even strong {\MgII} absorbers are associated with
galaxies with luminosities as small as $0.05L^*$, this image does not
allow us to usefully constrain the host galaxy properties for this or
any of the other {\MgII} systems along the line of sight.


\subsection{The HIRES Spectra}
\label{sec:hires}

HIRES/Keck \citep{vogt94} spectra at $R\sim45,000$
(FWHM$\sim6.6$~{\kms}) were obtained on 1995 July 4 and 5.  The
wavelength coverage is $3723$ to $6186$~{\AA}.  The typical
signal-to-noise ratio per resolution element is $50$.  The transitions
we study here are {\MgIIdblt} and {\FeII}~$\lambda 2600$.
\citet{cv01} describe the data reduction, continuum fitting, and Voigt
profile fitting.  We adopt the column densities and Doppler parameters
presented in \citet{anticip}.

\subsection{The STIS Spectra}
\label{sec:stis}

Archival STIS/HST \citep{kimble} spectra were used to study {\Lya},
{\Lyb}, {\Lyg}, {\Lyd}, {\rm Ly}\kern 0.1em$8$, {\CII} $\lambda 1334$,
{\SiII} $\lambda 1260$, {\SiIII} $\lambda 1206$, {\SiIVdblt},
{\NVdblt}, and {\OVIdblt}.  Two data sets were obtained with different
tilts of the E230M grating with an aperture of $0.2$ x $0.2$
($R=30,000$, FWHM$\sim10$~{\kms}). The first was obtained by Jannuzi
{\etal} (proposal identification 8312, wavelength range of 2303 to
3111~{\AA}) and the second by Burles {\etal} (proposal identification
7292, wavelength range from $1865$ to $2673$~{\AA}).  A third data set
was obtained by Burles {\etal} (wavelength coverage $1830$ to
$1870$~{\AA}) with the $52$ x $0.02$ aperture of the G230M grating
($R=10,000$, FWHM$\sim30$~{\kms}).  We used this spectrum to study the
higher order transitions of the Lyman series and the Lyman limit
break.  The reduction and calibration of all STIS data were performed
using the standard STIS pipeline
\citep{stis1}.  Continuum fitting was performed using the techniques
described in \citet{cv01}.  We co-added the E230M spectra in cases of
duplicate wavelength coverage to obtain higher signal-to-noise ratios
(per resolution element) ranging from $8$ to $30$.  We have verified
that wavelength calibration is consistent between the STIS and HIRES
datasets by comparing {\MgIIdblt}, {\SiII}~1260 and {\CII}~1335 in the
three single component, low ionization absorption systems found along
this same line of sight \citep{weak1634}.

\subsection{The FOS Spectra}
\label{sec:fos}
FOS/HST data, with lower resolution ($R=1300$) and wavelength coverage
$2225$~{\AA} to $3280$~{\AA} (G270H grating), were obtained as part of
the Quasar Absorption Line Key Project \citep{cat1,cat2,cat3}.  These
data were only used to study the {\CIVdblt} absorption, since all
other transitions had higher resolution coverage from STIS.

\section{The $z=1.0414$ System}
\label{sec:system}

\begin{figure*}
\vglue -0.5in
\figurenum{1}
\epsscale{1.0}
\plotone{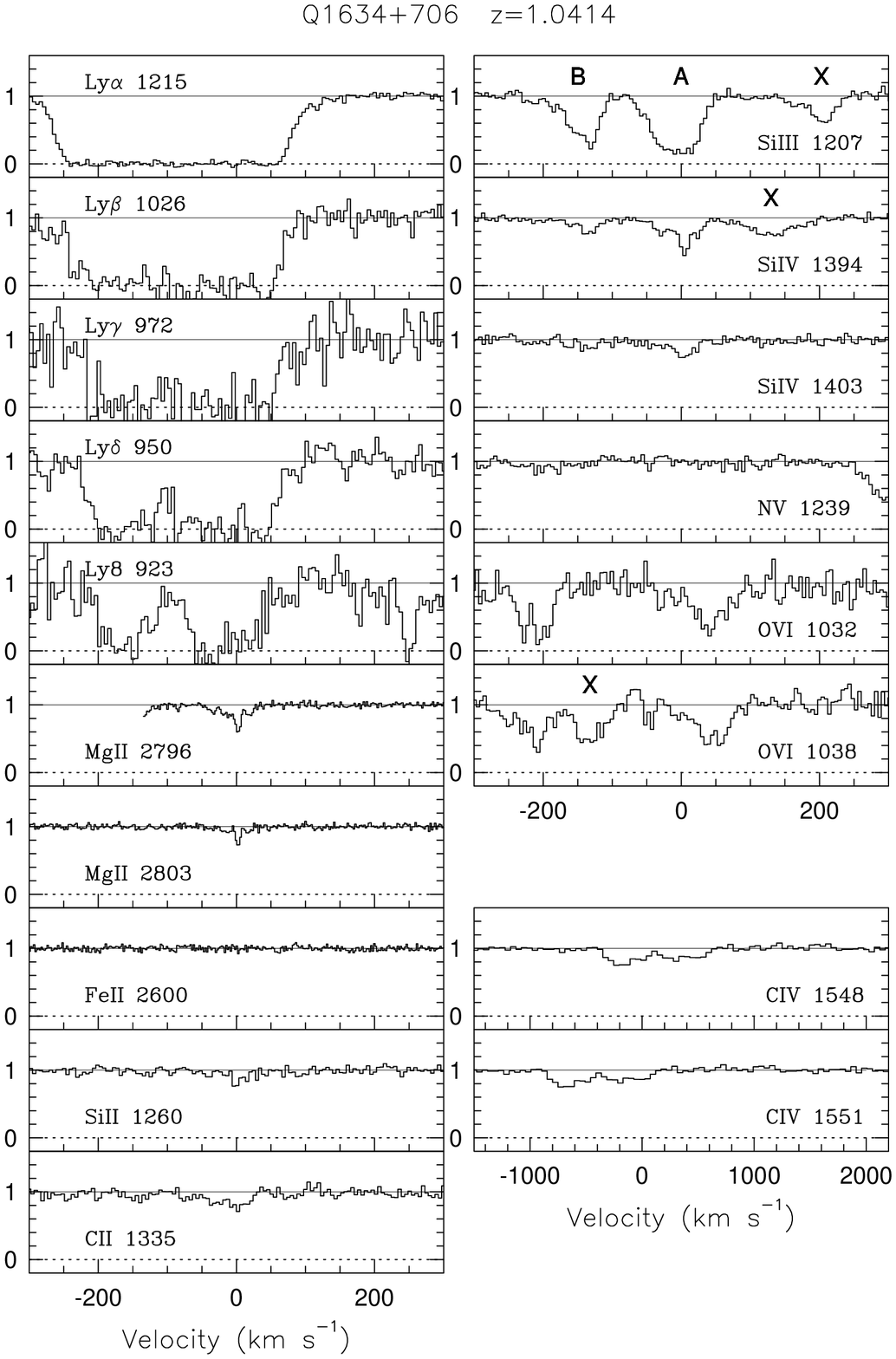}
\vglue -0.5in  
\protect\caption{
\baselineskip = 0.7\baselineskip
The high resolution absorption profiles for the $z=1.0414$ {\MgII}
absorber are presented in rest--frame velocity.  The {\MgIIdblt}
doublet and {\FeII}~$\lambda 2600$ were observed with HIRES/Keck
($R=45,000$); the {\CIVdblt} doublet was observed with FOS/HST
($R=1,300$); all other transitions were observed with STIS/{\it HST\/}
($R=30,000$).  Transitions not shown were either not detected or were
severely blended with other absorption features.  The {\MgII}~$\lambda
2796$ transition was not captured by the CCD for velocities less than
$-150$~{\kms}.  Two subsystems, at $v \simeq 0$ (subsystem A) and
$v~\simeq -150$~{\kms} (subsystem B), are apparent from the {\SiIII}~
$\lambda 1207$ transition as noted.  Contaminating absorption features
from systems not at $z=1.04$ are marked with an ``x'' above the
continuum.}
\label{fig:data}
\end{figure*}

Figure~\ref{fig:data} presents the absorption profiles of selected
transitions for the $z=1.04$ absorber.  These transitions are {\Lya},
{\MgIIdblt}, {\CII} $\lambda 1335$, {\SiII} $\lambda 1260$, {\SiIII}
$\lambda 1207$ {\SiIVdblt}, and {\OVIdblt}.  Also, {\CIVdblt}, covered
in the lower resolution FOS spectrum, is shown.  The region of the
spectrum that provides a limit on {\NV} $\lambda 1239$ and {\FeII}
$\lambda 2600$ is also shown.  The rest-frame velocities are aligned
with the {\MgII}~$\lambda 2796$ transition, where $v=0$ {\kms} is
defined at the apparent optical depth median of the profile.
Table~$1$ gives the rest frame equivalent widths or $3\sigma$
equivalent width limits for the illustrated transitions.

In the STIS spectra, two low--ionization ``kinematic subsystems'',
separated by $\sim150$~{\kms}, are detected in {\SiIII} and {\SiIV}.
We refer to these as subsystem A (at $\sim0$ {\kms}) and subsystem B
(at $\sim-150$ {\kms}). In addition, the {\OVI} doublet is detected in
two broad components, one $\sim50$~{\kms} to the red of subsystem A
(red broad component) and the other $\sim50$~{\kms} to the blue of
subsystem B (blue broad component).  Table~$1$ lists separately the
contributions to the equivalent width from subsystems A and B.

\begin{figure*}
\figurenum{2}
\epsscale{1.0}
\plotone{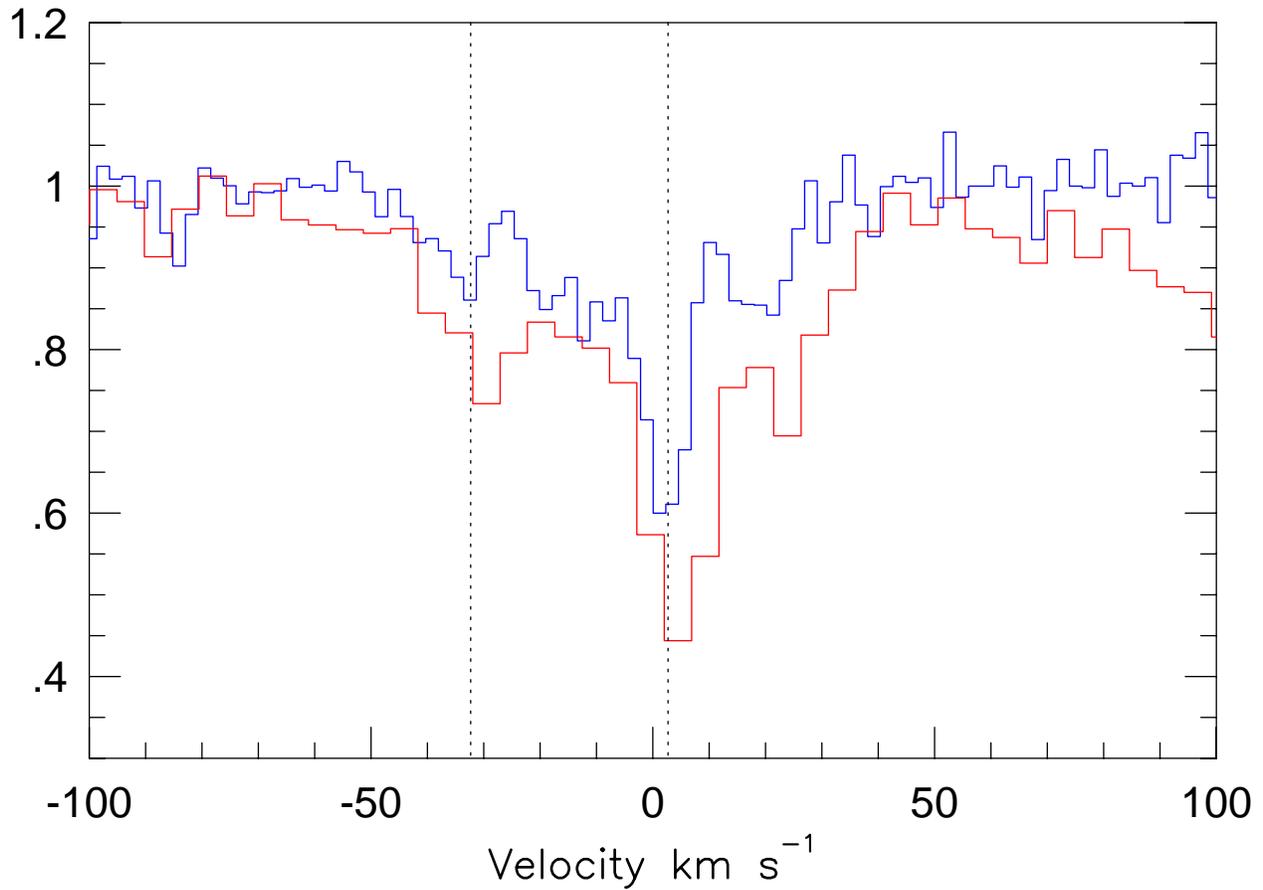}  
\protect\caption{
\baselineskip = 0.7\baselineskip
The {\MgII}~$\lambda 2796$ from HIRES/Keck and the {\SiIV}~$\lambda
1394$ profile from STIS/HST are superimposed in velocity space, the
weaker profile corresponding to {\MgII}.  Although {\MgII} and {\SiIV}
appears to have similar structure, they appear to be offset by a few
{\kms}.  Vertical lines are drawn through the apparent centroid of two
of the components in the {\MgII} profile.}
\label{fig:offset}
\end{figure*}

In subsystem A, the {\MgII} profile was fitted with four Voigt profile
components over a $50$~{\kms} interval \citep{anticip}.
The {\SiIV} shows the same kinematic structure, yet there may be
a very slight kinematic offset, with each of the four
components redward with velocity offsets ranging from
$1.3$--$5.1$~{\kms}.  To illustrate this, the {\MgII}~2796 and
{\SiIV}~1394 profiles are superimposed in Figure~\ref{fig:offset}.
As mentioned in section
\ref{sec:stis}, the alignment of similar transitions in other
absorbers along this same line of sight verifies that this is not
likely to be a result of wavelength calibration \citep{weak1634}.  

In contrast, the {\SiIII} profile is smooth.
In subsystem B ($v\sim-150$ {\kms}) an echelle order break prevented
observation of {\MgII}~$\lambda 2796$. However, the weaker
{\MgII}~$\lambda 2803$ transition is not detected.  The {\SiIII}
kinematics of subsystem B is characterized by an asymmetric blend with
a velocity width of $\sim 100$ {\kms}.

Figure~\ref{fig:data} also presents the {\CIVdblt} doublet observed with
FOS/HST.  The {\CIV} profiles are resolved, indicating a kinematic
velocity spread slightly greater than the instrumental resolution of
$230$~{\kms} (FWHM).   The {\CIV}
equivalent widths in Table~1 were derived from this lower
resolution FOS spectrum.
The upper panel of Figure~\ref{fig:lya} shows the
partial Lyman limit break observed with the G230M grating of STIS/HST.
The optical depth is $\tau \sim 1.3$, using the method of
\citet{dpsKP}, which implies $\log N({\HI})\sim 17.3$.


\section{Methodology for Modeling}
\label{sec:method}

We derive constraints on the kinematic, chemical, and ionization
conditions of the absorbing gas.  Our method is to compare the
absorption profiles of the observed chemical transitions (presented in
Figures~$1$ and $2$) with those predicted by photoionization and
collisional ionization models.  We summarize the method here, and it
is also described and applied in our earlier papers \citep{ding1634,
weak1634,ding1206}.

For photoionization, we use the code Cloudy \citep{ferland} and adopt
the extragalactic background spectrum of \citet{haardtmadau96} for the
ultraviolet ionizing flux.  We use the spectrum normalized at $z=1$.
We have also considered the influence of stellar ionizing flux, which
we describe below in \S~\ref{sec:stellar}.  For collisional ionization,
we adopt the equilibrium models of \citet{sd93}, which provide the
column densities of various transitions for assumed temperatures and
metallicities.

For photoionization, each cloud is modeled as a constant density
plane-parallel slab with the ionizing flux incident on one face.  We
assume a solar abundance pattern, however, deviations from this
pattern are considered when suggested by the data.  The ionization
conditions are defined by the ionization parameter, $\log U$, where
$U=n_{\gamma}/n_{\rm H}$, the ratio of the number density of hydrogen
ionizing photons to the number density of hydrogen.  The quantity
$n_{\gamma}$ is fixed by the normalization of the extragalactic
background spectrum, so that the relationship between $\log U$ and $n_{\rm
H}$ is $\log U= -5.2 - \log n_{\rm H}$.  Since the abundance pattern and
ionization parameter are interdependent, the models are not unique.

We have found that the absorption lines cannot be modeled as a single
ionization ``phase.''  In this context, phases denote regions or
``clouds'' with similar ionization parameter, density, metallicity,
and temperature.  For simplicity, clouds in the same phase are assumed
to have the same metallicity.  Moreover, we consider physical
scenarios that require a minimum number of phases to explain the data.

The lower ionization transitions in the $z=1.04$ absorber reveal clear
kinematic structure and we therefore use them as a template to place
constraints on the phase structure of the system.  In order to obtain
the best initial estimates for the cloud properties, we begin with the
lowest ionization transition that is most clearly detected and
resolved in each of the subsystems.  We obtain the column densities
and Doppler $b$ parameters for this transition using Voigt profile
(VP) fits.  In subsystem A, we adopt the column densities and Doppler
parameters for the {\MgII} doublet from \citet{anticip} as stated
above.  In subsystem B, {\MgII}~$2796$ is not covered by Keck and so
we fitted the {\SiIII}~$\lambda 1207$ profile.  VP profile fits are
dependent on spectral resolution.  Although they do not provide a
unique description of the physical conditions of the gas, we can still
use them to identify a plausible range of parameter space.  The
{\SiIII} data are of lower resolution than the {\MgII} data.  Compared
to the {\MgII}, the fits to the {\SiIII} grouping will in general
likely yield fewer clouds, which will have larger $b$ parameters and
smaller column densities, and this is likely to be a bias rather than
a real, physical effect.

We use the optimization mode of Cloudy.  For each cloud, we
``optimize'' on the VP column density of a selected transition, for an
assumed ionization parameter and metallicity.  Cloudy then calculates
the column densities for all other ionization species and determines
the kinetic temperature.  This temperature is used to calculate the
thermal component of the $b$ parameter for the optimized transition.
This thermal $b$ is then compared to the VP $b$ parameter to calculate
the turbulent component assuming Gaussian turbulence.  Doppler
parameters for all other transitions are then computed using this
turbulent component and scaling for ion mass.

To compare the Cloudy model predictions to the observations, we
generate synthetic spectra based upon the Cloudy column densities and
derived $b$ parameters.  The synthetic spectra are convolved with the
appropriate instrumental profile, and are assigned the same pixel
sampling as the instrument with which a given transition was observed.
For HIRES and FOS, we used Gaussians with FWHM$=6.6$~{\kms} and
$230$~{\kms}, respectively.  For STIS, the instrumental spread
function for the E230M grating was taken from the STIS performance
web pages at {\it www.stsci.edu}.

The ionization parameters and metallicities
of Cloudy models are then iteratively adjusted in order to minimize
deviations between the observed and synthetic spectra.  Although we
considered a strict $\chi^2$ criterion for optimization, we found
this to be unrealistic given the tendency for one or two pixels
to have a dominant contribution to that statistic.  Visual inspection,
comparing the model profiles to the data, generally yield ionization parameters
and metallicities accurate to $\sim 0.1$--$0.2$ dex.

The ionization conditions are determined by examining the relative
absorption strengths of the different metal transitions.  For the
optically thin regime, the entire cloud is subject to the same
ionizing spectrum (there is no ionization structure) and thus $\log U$
can be determined from two metal transitions independent of
metallicity.  This point is illustrated in Fig.~\ref{fig:cloudyill},
which shows two grids of Cloudy models with different metallicities,
$0.01$~solar and solar.  To construct the grids, the neutral
hydrogen column density was fixed at $\log N({\HI}) = 17$, and
the ionization parameter was varied from $\log U = -5$ to $\log U = -1$.
The figure shows that the ratio of the column densities of various
transitions, at a given $\log U$, does not depend on the metallicity.
For the allowed range of $\log U$, metallicity is then
constrained by the partial Lyman limit break and the {\Lya}
transition.

In this paper, we have assumed a solar abundance for the optimized
transition.  However, it is straightforward to consider how the
metallicity derived would differ if we assumed a different abundance.
If the optimized transition has an abundance of $A$~dex relative to
solar, then a model with a metallicity of $\log Z = \log Z_{orig} - A$
would be correct.  Of course, the abundances of other elements would
have to be adjusted relative to the optimized transition based on
whatever abundance pattern was assumed.


\section{Results}
\label{sec:results}

We present constraints on the gross properties of the $z=1.0414$
absorber along the line of sight to PG~$1634+706$.  We do not attempt
to find a unique model to describe the data, however we consider what
general conclusions can be drawn about the kinematics and multi-phase
ionization structure of the absorbing gas.  We present examples of
plausible models of the column densities, Doppler parameters,
ionization parameters, and metallicities in Tables $2$ and $3$.

\subsection{Low--Ionization Narrow Components}
\label{sec:narrow}

We constrained the range of ionization parameters for subsystems A and
B using detected metal transitions and limits.  For subsystem A, we
consider two models: one in which the {\MgII} and {\SiIV}
arise in the same phase (Model 1), and the other in which they arise in
different phases (Model 2).

\subsubsection{Ionization Conditions of Subsystem A - Model 1}

The VP fits yield four clouds with {\MgII} column densities ranging
from $\log N({\MgII})= 11.5$ to $12.0$ and $b({\rm Mg})$
parameters from $3$ to $11$~{\kms} (see Table $2$). In Model 1 we
assume that the {\SiIV} and {\MgII} arise in the same phase. To fully
account for the {\SiIV} absorption the four clouds would have $\log U$
values of $-2.3$, $-2.5$, $-2.3$, and $-2.3$, respectively.
Also, the abundances of silicon and carbon must be decreased
to make Model 1 consistent with the data.
With a solar abundance pattern, and
these $\log U$ values, {\SiII}, {\CII}, {\SiIII}, {\SiIV}, and
{\CIV} would be overproduced relative to {\MgII}.
A $\sim0.5$~dex decrease of silicon and carbon, relative to magnesium,
resolves this discrepancy.

Figures~\ref{fig:model1a} and \ref{fig:model1b} present this model
(Model 1), with and without the abundance pattern adjustment,
superimposed on the data.  The small kinematic offset between {\SiIV}
and {\MgII} (mentioned in \S~\ref{sec:system}) is apparent in the
{\SiIV}~$1394$ panel as well as in the expanded data in
Figure~\ref{fig:offset}.  This leads us to consider an alternative
model, in which the {\MgII} and {\SiIV} are produced by separate,
kinematically offset clouds.

\subsubsection{Ionization Conditions of Subsystem A - Model 2}
\label{sec:suba}

In Model 2 we assume that the {\SiIV} absorption for subsystem A
is produced entirely in a phase separate from the {\MgII}.
We again choose {\MgII} as the optimized transition for the
lower ionization ``{\MgII} clouds'', and use the VP fits to
the {\MgII} as the starting point for our model.  However, now
we also perform VP fits to the {\SiIV}, obtaining four additional
``{\SiIV} clouds'', and use {\SiIV} as the
optimized transition for this separate phase.

For the {\MgII} clouds,
insignificant {\SiIV} is produced at the same velocity if $\log U \la -3.0$.
The ionization conditions for the {\MgII} clouds can also
be constrained by {\FeII}~$\lambda 2600$.  {\FeII}~$\lambda 2600$ is not
detected, providing an upper limit of $\log N({\FeII}) <11.5$.
\footnote{\baselineskip = 0.5\baselineskip
This column density is derived from the $3$ $\sigma$ limit of the
equivalent width assuming a Doppler parameter of $10$~{\kms}.  The
curve of growth of {\FeII}~$\lambda 2600$ for this equivalent width is
linear and therefore independent of Doppler parameter.
The $N({\FeII})$ limit is derived at the position of each {\MgII}
component, assuming an unresolved {\FeII} line at that velocity.}
In principle,
this would provide a lower limit on $\log U$. However, below $\log U =
-4.0$, $N({\FeII})/N({\MgII})$ depends only weakly on $\log U$.
Significant {\FeII} absorption is not produced, even in models
with $\log U$ as low as $-6$.  As with Model 1, these {\MgII} clouds
overproduce {\SiII} and {\CII} over the full range of $\log U$.  A
similar abundance pattern adjustment ($\sim -0.5$~dex) of carbon and
silicon is required.  The properties for a sample version of Model 2
for these {\MgII} clouds, with $\log U = -3.1$, near its maximum, are
given in Table~2.  The contribution of these {\MgII} clouds to the
absorption in other transitions (without the $0.5$~dex abundance
pattern adjustment) is show as the dotted curve in
Figures~\ref{fig:model2a} and \ref{fig:model2b}.

To produce the {\SiIV} absorption in Model 2, we assumed each
individual component was shifted by a slightly different velocity
(given in Table~2).  These additional clouds, centered on the {\SiIV}
(``{\SiIV} clouds''),
would have $ -2.9 < \log U < -1.9$.  A lower $\log U$ would produce
{\MgII} in these clouds as well (thus overproducing {\MgII} in the
combined {\MgII} and {\SiIV} clouds) and a higher $\log U$ produces
{\OVI}, which is not detected at this velocity.

This model including the {\MgII} and {\SiIV} clouds is superimposed as
a long--dashed line on the data in Figures~\ref{fig:model2a} and
\ref{fig:model2b}, respectively.  The agreement with the observed
{\SiIV}~$1394$ profile is better than in Model 1 (see
Figure~\ref{fig:model1b}).  However, in both cases, the model {\SiIII}
profile, which is produced mainly by the clouds centered on {\SiIV},
has a definite cloud structure which is not apparent in the data.
This indicates that neither model is fully consistent with the
observed {\SiIII} absorption and suggests that an additional phase is
needed to fit the {\SiIII}.  Also, the strongest {\SiIV} cloud
overproduces {\SiIII} at $3.9$ {\kms}.  The ionization parameter
cannot be adjusted to fit both the {\SiIII} and {\SiIV} without
overproducing higher ionization transitions, like {\OVI}.  If the
ionization parameter is increased to fit these two transitions, oxygen
would have to be deficient with respect to silicon.

We also considered a model in which both the {\SiIV} clouds and {\MgII}
clouds would contribute to the {\SiIII} profile.  Since the {\SiIV} and
{\MgII} clouds are ``staggered'' in velocity, they
could then, in principle, blend out the structure within
the {\SiIII} and produce the ``flat-bottom'' profile seen in the data.
However, the combined contribution to {\SiIII} is too large and
{\SiIII} is overproduced.  The ionization parameters cannot be
adjusted in such a way as to decrease the total contribution to
{\SiIII} without overproducing {\OVI}.  This alternative
model may be excluded.

\subsubsection{Ionization Conditions of Subsystem B}
\label{sec:subb}

VP fits to the {\SiIII} profile (at $-200 < v < -100$~{\kms}) yield
three clouds (``{\SiIII} clouds'') with a range of column densities
from $\log N=11.9$ to $12.8$ and $b$ parameters from $6$ to
$22$~{\kms} (see Table~$2$).  {\MgII} $\lambda 2803$ and {\OVI} were
not detected in this velocity range.  The limits on these transitions
were used to constrain the ionization conditions of the {\SiIII}
clouds, at $v=-145$~{\kms} and $v=-128$~{\kms}.  The properties of the
cloud at $v=-185$~{\kms}, which is required to fit the {\SiIII}, are
not well constrained because it is weak and not definitively detected
in other transitions.  For this reason, we assume it has the same
ionization parameter as the other clouds in subsystem B.  A lower
limit of $\log U\simeq -3.0$ is derived from the lack of {\MgII}
$\lambda 2803$ absorption.  An upper limit of $\log U\simeq -2.0$
applies in order that {\OVI} is not overproduced.  A model with $\log
U = -2.7$ for the subsystem B clouds is superimposed on the data in
Figures~\ref{fig:model1a},
\ref{fig:model1b}, \ref{fig:model2a} and
\ref{fig:model2b}.

\subsubsection{Metallicity}
\label{sec:metals}

\begin{figure*}
\figurenum{3}
\epsscale{0.85}
\plotone{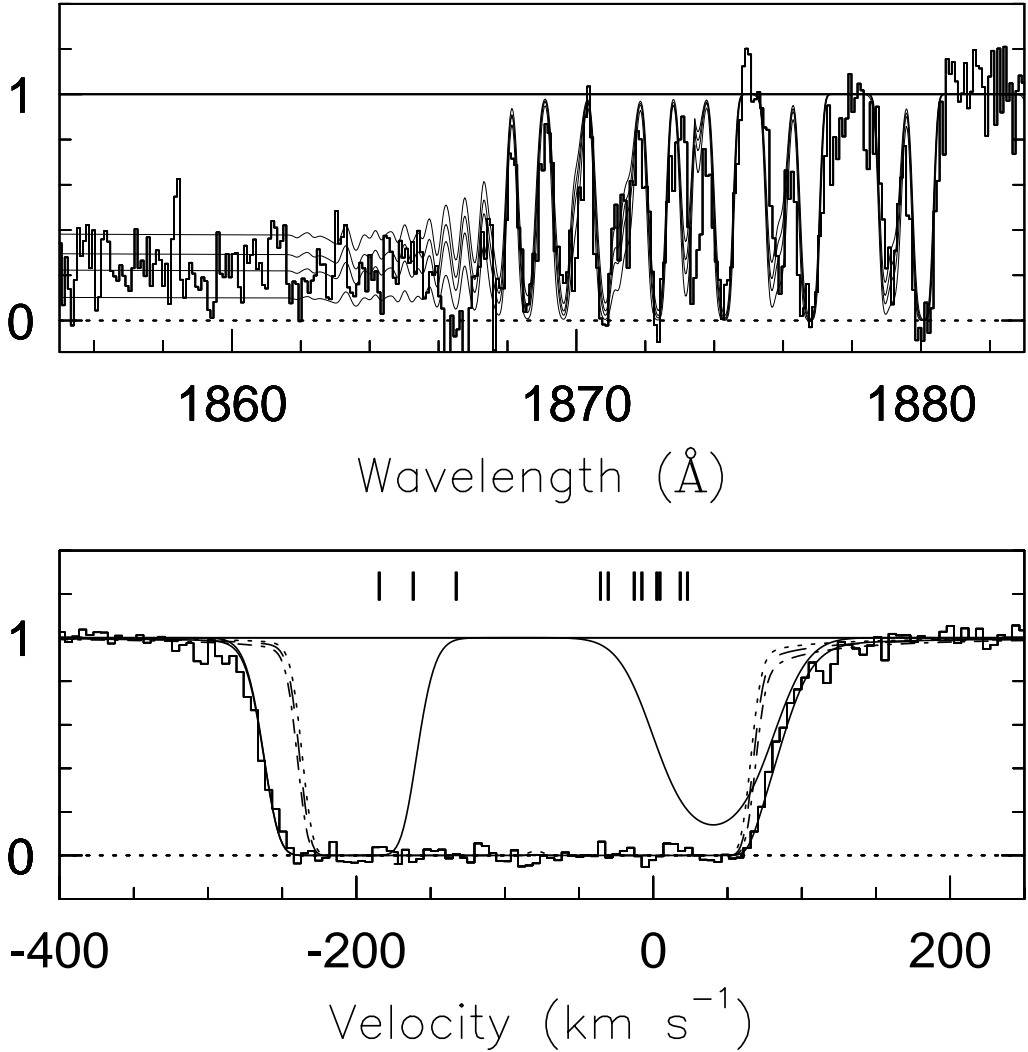}  
\protect\caption{
\baselineskip = 0.7\baselineskip
(a) The partial Lyman limit break, captured in the G230M/STIS
spectrum, provided the best constraint on the metallicity of the two
subsystems.  For simplicity, all clouds in both subsystems were
assumed to have the same metallicity.  Shown in this figure are
synthesized spectra for models (using Model 2 for subsystem A) with
metallicities of $\log Z = -1.3$ (upper curve), $\log Z = -1.4$ (our
best fit), $\log Z = -1.5$, and $\log Z = -1.7$ (lower curve).  The
broad components make a negligible contribution to the Lyman limit
break.  These metallicities correspond to {\HI} column densities of
$\log N({\HI}) = 17.8, 17.23, 17.38$ and $17.56$ respectively.  The
saturated absorption feature just blueward of the Lyman limit break is
due to {\HI} $\lambda 938$ from the strong {\MgII} absorber at
$z=0.9902$ ---(b) The {\Lya} absorption line was covered by the
E230M/STIS spectrum.  Synthesized spectra for model contribution from
the subsystems, with metallicities of $\log Z = -1.3$, $\log Z =
-1.4$, $\log Z = -1.5$, and $\log Z = -1.7$ are overplotted (dashed
lines) on these data.  Higher metallicities produce smaller absorption
strengths.  Note that the absorption is not fully produced by the
subsystems in either the red or the blue wing.  The lines above this
plot mark the position of the subsystem clouds in model 2.  Although
model 2 and model 1 differ in their source of {\SiIV} production, the
{\Lya} profile is nearly unaffected by this effect.  The three solid
lines show the contributions from the two broad components and the
full model, combining all componenets.  Parameters defining our ``best
fit'' model, including the two broad components as well as the
subsystems can be found in Table 2 (Model 2) and Table 3.}
\label{fig:lya}
\end{figure*}

\begin{figure*}
\figurenum{4}
\epsscale{1.0}
\plotone{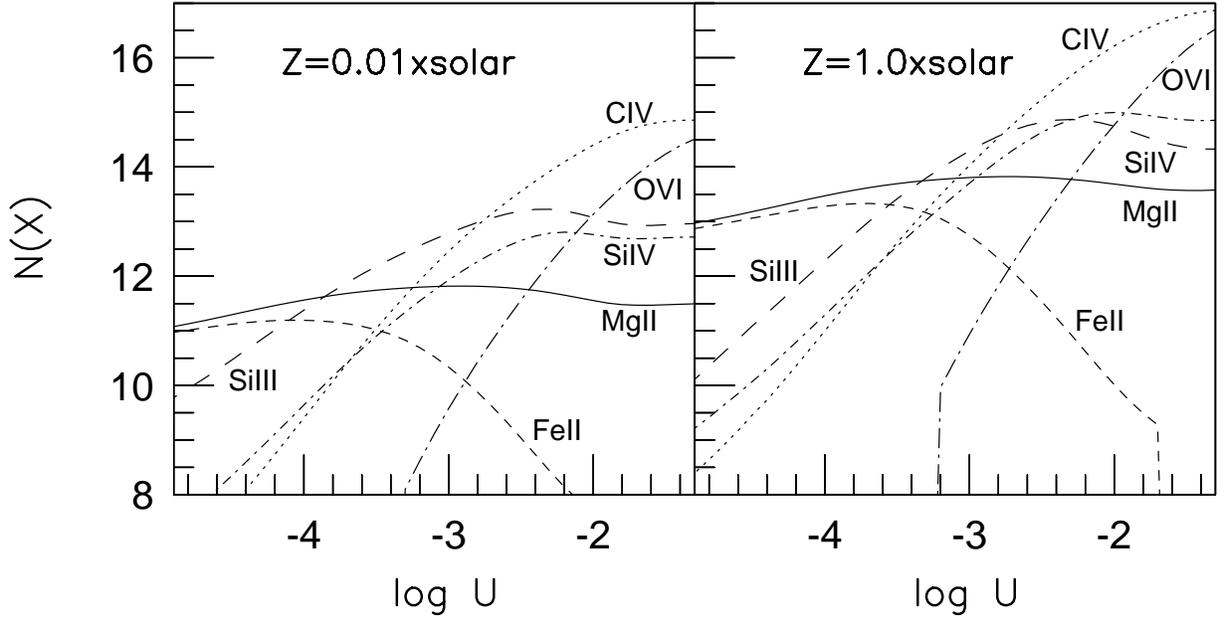}  
\protect\caption{
\baselineskip = 0.7\baselineskip
The column densities of selected transitions as a function of
ionization parameter, for Cloudy photoionization models with $\log
N({\HI}) = 17$.  The different curves represent column densities of
various transitions.  The left panel shows the results for metallicity
$0.01\times$solar and the right panel for a solar metallicity.  The
ratios of the different transitions is relatively constant,
independent of metallicity, although for metallicities solar or higher
there are small differences from the results for lower metallicities.
Because of the constant ratios, the ionization parameter can be
constrained, independent of metallicity.  Afterwards, the
``normalization'' can be set by adjusting the metallicity (adjusting
$N({\HI})$) to fit {\Lya} and/or the Lyman limit break.  The curve
representing {\OVI} has a sharp ``cut-off'' feature at $log U = -3.2$.
This is not a real effect but rather a result of the limitation of
Cloudy to evaluate the column density of {\OVI} in this regime.}
\label{fig:cloudyill}
\end{figure*}

\begin{figure*}
\vglue -1.0in
\figurenum{5}
\epsscale{1.1}
\plotone{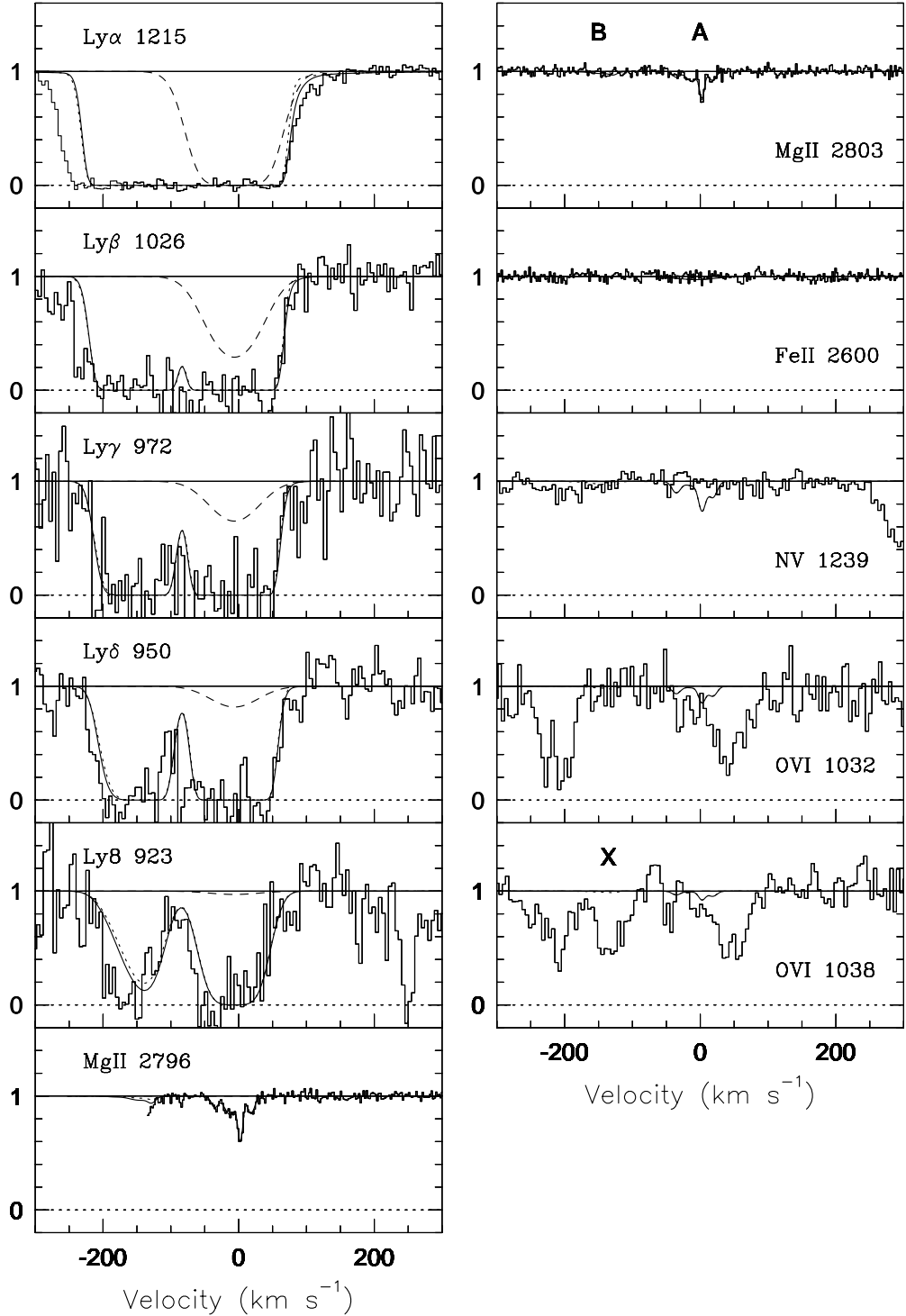}
\vglue -0.9in
\protect\caption{
\baselineskip = 0.7\baselineskip
Spectra covering various transitions are shown in velocity space, as
in Figure~\ref{fig:data}, with contributions from models of the
subsystems superimposed.  Model 1 for subsystem A (see Table~$2$) is
illustrated, in which the {\SiIV} is produced in the same phase as the
{\MgII}.  The dotted lines represent model contributions from the
{\MgII} clouds for subsystem A when a solar abundance pattern is
assumed.  The solid model curves have the abundances of silicon and
carbon decreased by $0.5$~dex relative to magnesium.  The dashed
curves represent the absorption contribution (mostly to {\SiIII}) from
a collisionally ionized component with $\log T = 4.62$.  The solid
curve superimposed on subsystem B represents the best fitting model
for the {\SiIII} clouds.  For this subsystem no abundance pattern
adjustments are required.  More transitions are included in
Figure~\ref{fig:model1b}.}
\label{fig:model1a}
\end{figure*}

For the range of $\log U$ derived in sections \ref{sec:suba} and
\ref{sec:subb}, we consider the constraints on metallicity for
subsystems A and B.  Since the column densities of the {\MgII} and
{\SiIII} transitions remain approximately constant from $-3<\log U
<-2$ (within $0.5$ dex: see Figure \ref{fig:cloudyill}), the metallicity
constraint is only weakly dependent on the $\log U$. The metallicity
of the subsystems is constrained by the partial Lyman limit break.
For simplicity, we assume that all the clouds in both subsystems, A
and B, have the same metallicity.  Figure~\ref{fig:lya}a shows
synthesized spectra (using Model 2 for subsystem A), for four
metallicities, expressed in solar units, ($\log Z = -1.3, -1.4, -1.5,
-1.7$) superimposed on the observed partial Lyman limit break. We find
that $\log Z$ is clearly between $\sim-1.3$ and $-1.7$.  For $\log Z >
-1.3$, $N({\HI})$ values would be too small to account for the
observed partial Lyman limit break.  Although the break could arise
from a separate phase, the metal transitions do not suggest such a
phase.  A lower metallicity is not possible because a larger
$N({\HI})$ would leave too little flux beyond the break.

Models 1 and 2 for subsystem A yield the same final result for
metallicity, with a best fit value of $\log Z = -1.4$.  For the
allowed range of metallicities, the break would have significant
contributions from several different clouds with $N({\HI})\sim 16$ to
$16.5$~{\cmsq} from both subsystem A and subsystem B.  Alternatively,
one of these clouds could have a lower metallicity so that it would
fully account for the partial Lyman limit break.  To compensate, the
other clouds would then have higher metallicities.

Figure~\ref{fig:lya}b shows that the {\Lya} absorption produced by our
models for the two subsystems, with $\log Z = -1.4$ does not fully
account for the absorption in the wings of the {\Lya} seen in the
data.  This implies that additional phases are needed, as will be
described in \S~\ref{sec:broad}.

\subsubsection{Cloud Sizes}
\label{sec:sizes}

Knowing the inferred ionization parameters and metallicities, the
cloud sizes can be estimated.  The size of a cloud scales with the
total hydrogen column density: $s = N({\rm H}_{tot})/n$, where
$n=10^{(-5.2-\log U)}$, assuming a Haardt-Madau background ionizing
spectrum.  Therefore, the cloud size scales inversely with
metallicity.  It is important to note that the ``size'' corresponds to
the thickness of the slab assumed in the photoionization models.
Realistically, a variety of geometries are possible.

For subsystem A, the two models that we have presented produce very
different conclusions about the cloud sizes.  In Model 1, a large
ionization parameter ($\log U \sim -2.3$) is required in order that
the {\SiIV} be produced in the same phase with the {\MgII}.
Large cloud sizes, ranging from $6$--$24$~kpc, result for these four
{\MgII} clouds.  In contrast, Model 2 separates the total
column density into two different phases, the {\MgII} and {\SiIV}
clouds, which have ionization parameters of $< -3.0$ and $\sim -2.3$,
respectively.  The {\MgII} clouds would be relatively small, with
upper limits on their sizes of $\sim100$--$300$~pc, while the {\SiIV}
clouds would be larger, $2$--$5$~kpc.
For subsystem B, the three clouds would have sizes of $80$--$800$~pc.


\subsubsection{Collisional Layer in Subsystem A}
\label{sec:collisional}

The Cloudy photoionization modeling process takes into account
collisional ionization based on the temperature derived for a system
in equilibrium. However, a separate layer not in equilibrium, e.g. due
to a shock, could exist within a system providing an additional
heating effect.  In this additional layer, collisional ionization
could dominate, giving rise to different absorption properties.

\begin{figure*}
\figurenum{6}
\epsscale{1.0}
\plotone{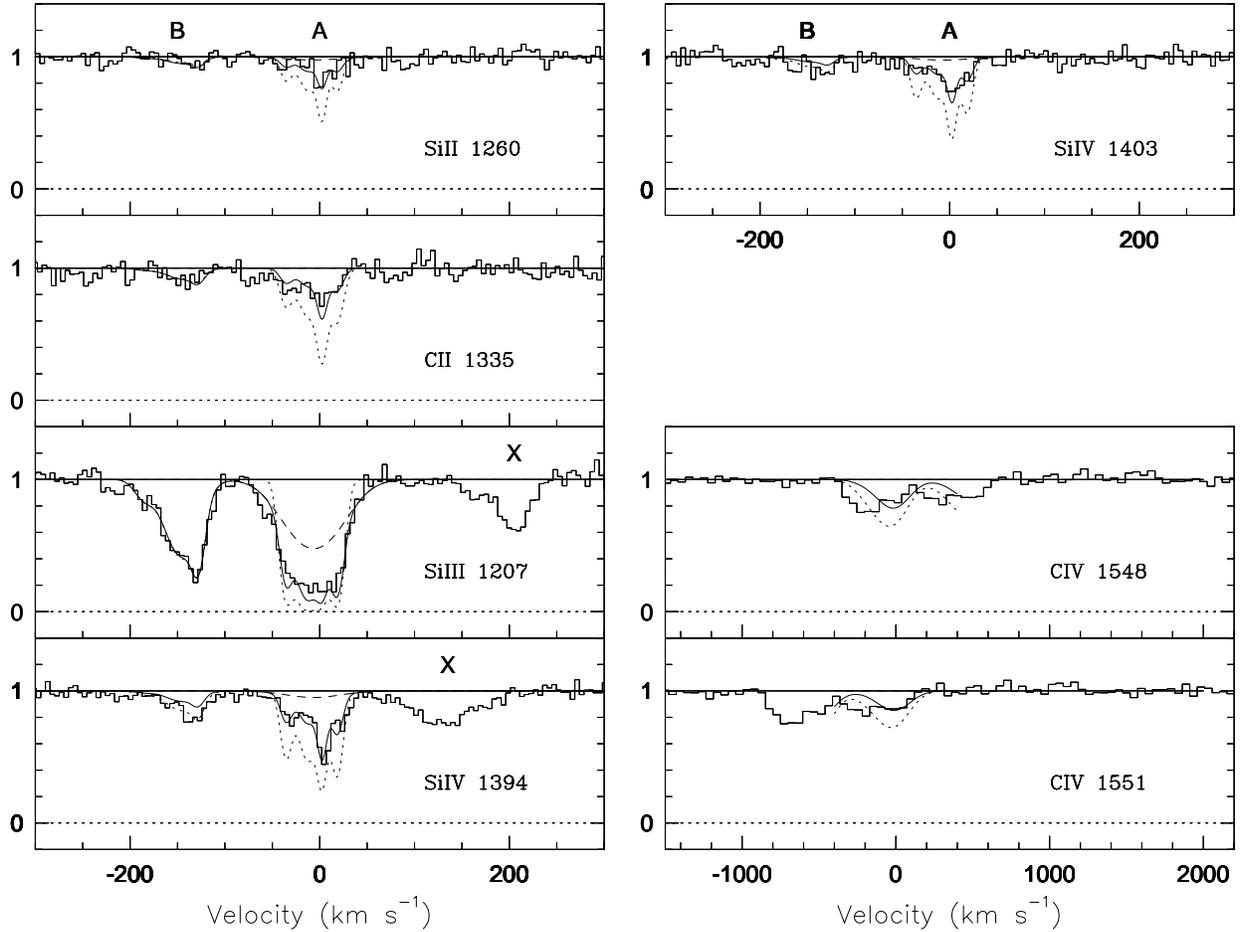}
\protect\caption{
\baselineskip = 0.7\baselineskip
As in Figure~\ref{fig:model1a}, synthesized spectra corresponding to
model 1 are overplotted on the data for additional transitions.  The
dotted line depicts the contribution from the {\MgII} clouds when a
solar abundace is applied.  Clearly, this overpredicts each of the
transitions presented in this figure.  An abundance pattern adjustment
in which carbon and silicon are deficient with respect to magnesium by
$0.5$~dex is overplotted as a solid line.  }
\label{fig:model1b}
\end{figure*}

\begin{figure*}
\vglue -1.0in
\figurenum{7}
\epsscale{1.1}
\plotone{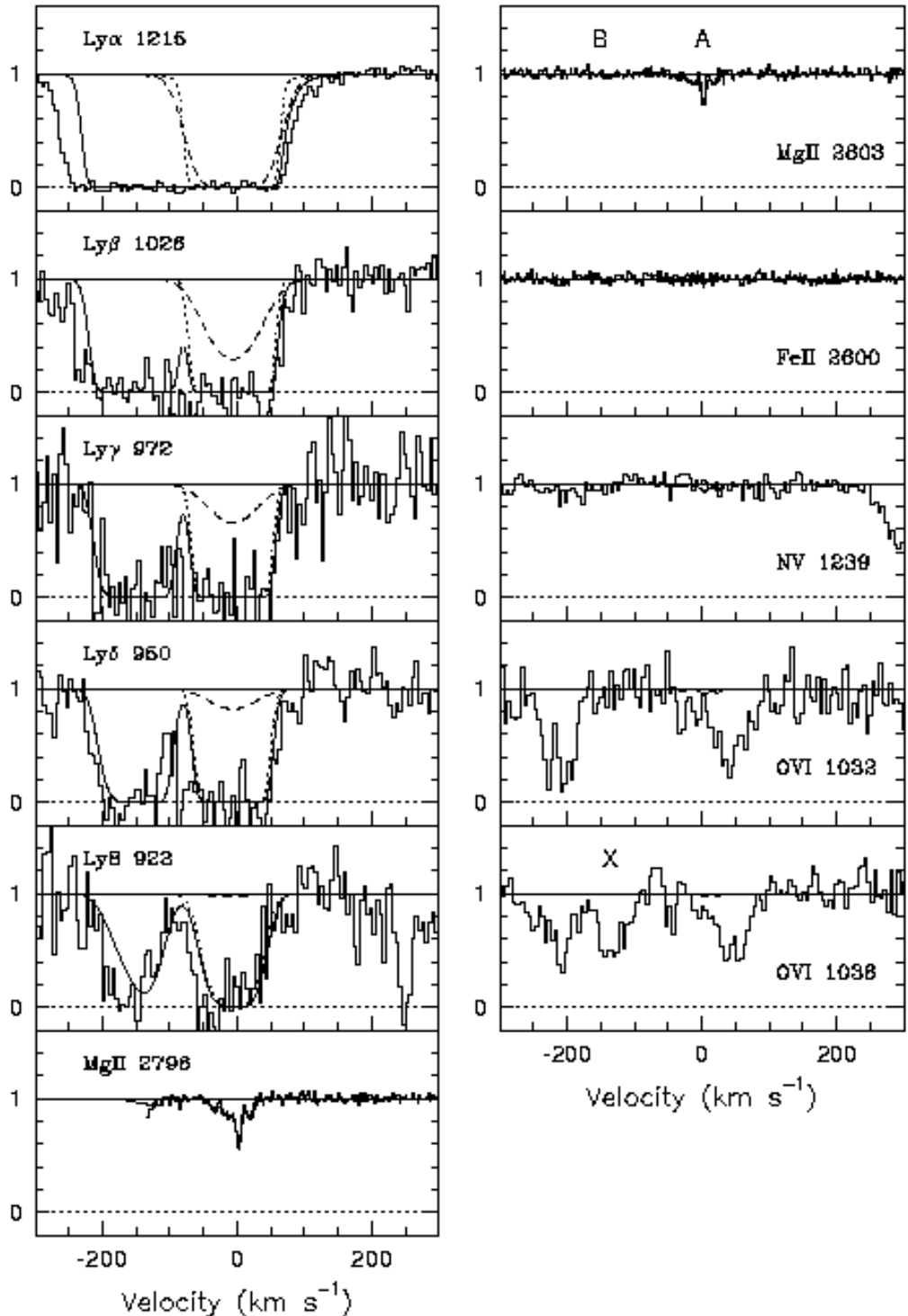}
\vglue -0.9in
\protect\caption{
\baselineskip = 0.7\baselineskip
Model contributions to the subsystems, as in Figure~\ref{fig:model1a},
but with Model 2 used to represent subsystem A (see Table~$2$).  The
dotted curves, representing the absorption contributions from the
{\MgII} clouds in subsystem A, now make a negligible contribution to
the {\SiIV} absorption.  The collisional component with $\log T =
4.62$ is shown as a dashed curve.  The long--dashed curves include
contributions from both the {\MgII} clouds and the additional offset
{\SiIV} clouds (using a solar abundance pattern for both), as well as
from the collisional component.  The solid curve includes all of these
contributions, but with a $0.5$~dex decrease in the abundances of
carbon and silicon in the {\MgII} clouds of subsystem A.  The
subsystem B model is the same as for Figure~\ref{fig:model1a}, again
with no abundance pattern adjustment.  Additional transitions are
shown in Figure~\ref{fig:model2b}.  }
\label{fig:model2a}
\end{figure*}

\begin{figure*}
\figurenum{8}
\epsscale{1.0}
\plotone{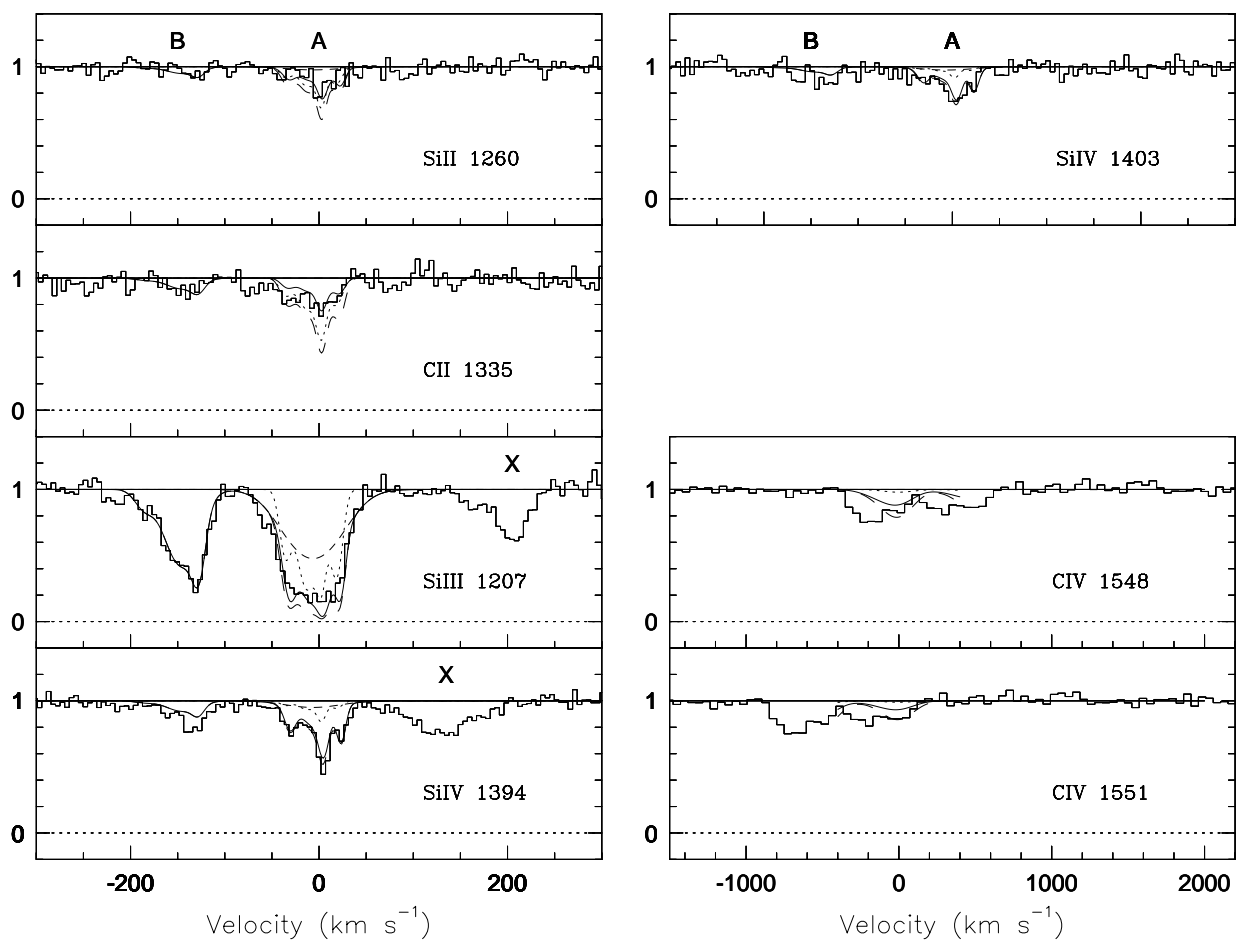}
\protect\caption{
\baselineskip = 0.7\baselineskip
As in Figure \ref{fig:model2a}, synthesized spectra corresponding to
model 2 are overplotted on the data for additional transitions.  This
model apperas to be a better fit than compared to model 1, shown in
Figure \ref{fig:model1b}.  The dotted lines shown the contribution to
the absorption from the {\MgII} clouds in subsystem A.  The dashed
line reveals the contribution from the collisionally ionized
componenent that creates the smooth profile in the {\SiIII} $\lambda
1207$ transition.  The long-dashed line combines the four {\MgII} and
four {\SiIV} clouds as well as the collsionally ionized component with
solar metallicity. The solid line shows our best fit to the data using
model 2.}
\label{fig:model2b}
\end{figure*}

The combination of the subsystem A clouds (either the four in Model 1
or the eight in Model 2) produces distinct cloud structure in the
{\SiIII} and does not adequately fit its smooth profile.  This
discrepancy can be seen from the dotted curves in
Figures~\ref{fig:model1b} and \ref{fig:model2b}.  Photoionization
models also do not account for the ``wings'' of the observed {\SiIII}
absorption.  A collisionally ionized phase, however, would naturally
produce a broader, smoother profile of {\SiIII}.  {\SiIII} absorption
peaks at a temperature of $\log T\sim4.5$
\citep{sd93}. However, at this temperature, {\FeII}~$\lambda 2600$ is
overproduced.

A temperature of $\log T\sim4.62$ could produce {\SiIII} without
producing detectable {\FeII} and {\OVI}.  The metallicity of the layer
would have to be $\log Z > -0.5$ in order not to overproduce {\Lya}.
The absorption contribution from a model of a collisionally ionized
layer with these properties is given as a dashed line in
Figures~\ref{fig:model1a}, \ref{fig:model1b}, \ref{fig:model2a} and
\ref{fig:model2b}.

\subsection{High--Ionization Broad Components}
\label{sec:broad}

The need for a broad component or components is apparent from the
underproduction of
{\Lya} absorption by the narrow subsystem clouds (see
Figure~\ref{fig:lya}b and \S~\ref{sec:metals}).  Also, although two
{\OVI} features are apparent, their velocity offsets ($50$ {\kms} for
both) from the subsystems makes it impossible that they are produced
in the same phase of gas.  Although {\CIV} is only covered at low
resolution, it is clear that the blue part of the profile is not
produced by the low--ionization components from the subsystems
discussed in
\S~\ref{sec:narrow}.  The red part of the {\CIV} profile is fully
produced by the subsystems, both for Model 1 and Model 2 of subsystem
A (see the solid curves in Figures~\ref{fig:model1b} and
\ref{fig:model2b}), with Model 1 producing only slightly more {\CIV}.
Therefore, the considerations for the physical conditions of the broad
components are similar for both models.  The discussion that follows
pertains specifically to Model 2.

Two higher ionization phase components can be added to simultaneously
produce the needed {\Lya} and {\OVI} absorption.  That is, if these
components are centered on the two {\OVI} features, it is also
possible to fit the two extremes of the {\Lya} profile.

The fit to the {\OVI} absorption is uncertain, especially in the blue
broad component.  There are apparent inconsistencies between the
$\lambda 1032$ and $\lambda 1038$ transitions indicating that blends
are present. The data may also suggest that there is substructure and
these {\OVI} phases may actually contain multiple clouds.  In
particular, there may be additional narrower components at $-210$
{\kms} and $35$ {\kms}.  However, for simplicity, we model each broad
component as one cloud and optimize on {\Lya}.  From our fit to
{\Lya}, we determine that $\log N({\HI})\sim15.0$ and $b\sim27$~{\kms}
for the blue broad component and $\log N({\HI})\sim14.0$ and
$b\sim37$~{\kms} for the red broad component.

We consider collisional ionization and photoionization for these two
broad components.  The measured $b$ parameter for hydrogen from the
blue broad component ($27$ {\kms}) implies a maximum temperature, $T <
m b^2/2k$, of $\log T\sim 4.6$.  This is too cool for this broad
component to give rise to significant {\OVI} absorption through
collisional ionization. The measured $b$ for {\OVI} ($\sim20$ {\kms})
yields a larger maximum temperature of $\log T\sim 5.6$.  Although
this value does imply that {\OVI} could be produced by collisional
ionization, there would then have to be another phase that accounts
for the blue broad component of {\Lya}.

Similarly, the $b$ parameter of hydrogen for the red broad component
($37$ {\kms}) implies a maximum temperature of $\log T\sim4.9$, which,
again, is too low for {\OVI} to be produced through collisional
ionization.  The $b$ parameter of {\OVI} ($\sim26$ {\kms}) allows for
a maximum temperature of $\log T\sim5.8$.  At this temperature, most
of the oxygen would be in the form of {\OVII}. If some of
the broadening was due to turbulent motion, the actual temperature
would be lower and significant {\OVI} could be produced through
collisional ionization.  However, like the blue broad component, the
red broad component of {\Lya} would require an additional phase if the
{\OVI} were to be produced by this mechanism.

Photoionization appears to be a simpler solution.  Photoionized broad
components can be added in such a way as to produce {\OVI} while
simultaneously ``filling out'' the {\Lya} profile.  For a given
metallicity, if it is high enough, a $\log U$ can be chosen to
simultaneously produce the observed {\OVI} and {\Lya}.  For this
$\log U$ we can then consider whether the model is also consistent
with the observed {\CIV}.  This requires substantial {\CIV}
production in the blue broad component, but negligible {\CIV}
production in the red broad component.

First, we consider whether the broad components could have a
metallicity of $\log Z = -1.4$, the same as the low--ionization
subsystems.  For the blue broad component a model with $\log U =-1.2$
is consistent with the observed {\OVI} profiles, though there is
some uncertainty because of the apparent inconsistency between
{\OVI}~$1032$ and {\OVI}~$1038$.  However, the {\CIV} at this
velocity is underproduced, and in fact cannot be produced at this
metallicity for any ionization parameter.  Although the {\CIV}
could be produced in a different phase from the {\OVI}, for simplicity
we favor a higher metallicity model which has the minimum number
of phases.  For the red broad component, with $\log Z = -1.4$,
the maximum $N({\OVI})$ is produced for $-0.5 < \log U < -0.2$,
however this is not enough {\OVI} to match the observed profiles.
Therefore, for both the blue and the red broad components,
a larger metallicity is favored.

\begin{figure*}
\figurenum{9}
\epsscale{1.0}
\plotone{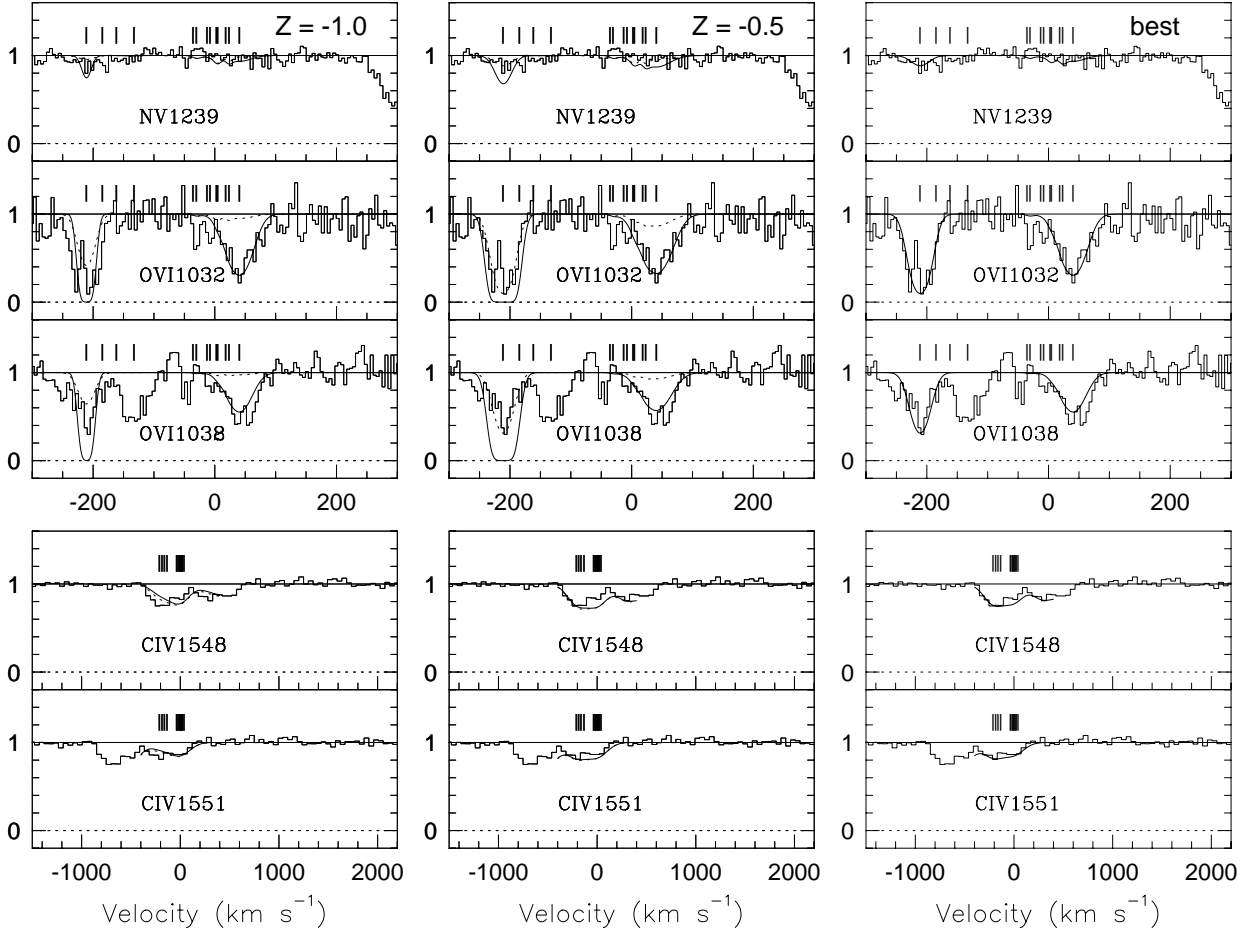}
\protect\caption{
\baselineskip = 0.7\baselineskip
The left panel illustrates, for a metallicity of $\log Z = -1.0$, the
effect of changing the ionization parameter of the two broad
components that produce high ionization absorption.  The dotted curves
represent a model with an ionization parameter of $\log U = -1.5$, and
the solid curves represent a model with $\log U = -0.5$.  For both
models, the abundance of nitrogen in the blue broad components has
been reduced by $0.8$~dex.  The subsystem contributions (important for
{\CIV} in subsystem A) are also included (using Model 2 for subsystem
A).  The middle panel is similar, but for $\log Z = -0.5$, and for
$\log U = -1.5$ (dotted) and $\log U = -1.0$ (solid), with a $0.8$~dex
reduction of nitrogen in the blue broad component.  The right hand
panel shows an acceptable model for the broad phases, with $\log Z =
-0.5$ and $\log U = -1.5$ for the blue broad component and $\log Z =
-1.0$ with $\log U = -0.5$ for the red broad component (see
Table~$3$).  This model includes a reduction of the abundance of
nitrogen by $\sim0.8$~dex.  }
\label{fig:broad}
\end{figure*}

Next, we consider $\log Z = -1.0$.  Models with this metallicity and
with various ionization parameters are superimposed on the high
ionization transition profiles in the left panel of Figure~\ref{fig:broad}.  For the
blue broad component, the model is consistent with the observed {\OVI}
profiles for $-1.4 < \log U < -1.2$, depending on the contribution of
blends to the two {\OVI} profiles.  However, to produce the observed
{\CIV} absorption, a somewhat lower $\log U \sim -1.8$ model is
needed.  We conclude that either a model with this metallicity is not
consistent with the observed blue broad component or that an abundance
pattern adjustment of carbon relative to oxygen applies.  The {\NV} is
also strongly overproduced by this model, so it requires an
abundance pattern adjustment of $\sim -0.8$~dex.  A model with $\log Z
= -1.0$ and $\log U = -1.4$, that achieves the best agreement with
{\OVI} and {\CIV}, produces a cloud with size $24$~kpc.  For the red
broad component, a $\log U = -0.5$ model produces an excellent fit to
the {\OVI} profiles (except for the apparent blend in the {\OVI}~1038
profile at $v\sim60$~{\kms}).  As required, it does not contribute
significantly to the {\CIV} absorption.  The cloud size for this model is quite
large, $257$~kpc.

Models with $\log Z = -0.5$ are also presented in
Figure~\ref{fig:broad}.  With a $\log Z = -0.5$ model, a single $\log
U$ value produces both the {\OVI} and {\CIV} in the blue broad
component, without adjustment of the abundance pattern.  The observed
{\OVI} and {\CIV} for the blue broad component can be produced by a
model with $\log Z = -0.5$ and $\log U = -1.5$.  Such a cloud has a
size of $12$~kpc.  Again, {\NV} is overproduced, and a reduction of
the nitrogen abundance by $0.8$~dex is required.  However, the red
broad component is overproduced at this metallicity and an ionization
parameter cannot be chosen in order to fit both the {\OVI} and {\CIV}
simultaneously.  For $\log U = -1$, which provides the best fit to
{\OVI}, but overproduces {\CIV}, the cloud would have a size of
$14$~kpc.

Because it is only captured in the lower--resolution
FOS data, the fit to the {\CIV} is quite uncertain.  For some
models, it appears that the {\CIV} $1548$ transition is overproduced,
yet the $1551$ transition is consistent.  For models with $\log Z =
-0.5$, both the $1548$ and the $1551$ absorption are overproduced in
the red broad component.  One possible resolution to this problem is
an abundance pattern adjustment in the red broad component, such that
carbon is deficient with respect to oxygen.  Alternatively, since
the {\SiIV} clouds from subsystem A also contribute
significantly to the {\CIV} at that velocity,
a model in which carbon is
deficient with respect to silicon in those clouds could also make a red broad
component with $\log Z = -0.5$ consistent with the observed data.

Another possibility is a scenario in which the broad components have
different metallicities.  Such a model is listed in Table~3, since it
has the fewer abundance pattern adjustments than the other consistent
models.  In this model, the blue broad component has
a metallicity of $\log Z = -0.5$, while the
red broad component has a metallicity of $\log Z = -1.0$ and a $\log
U = -0.5$.

\subsection{Effects of Spectral Shape}
\label{sec:stellar}

For all the modeling in the previous sections, we assumed a pure
Haardt and Madau extragalactic background photoionizing spectrum.
This is appropriate for absorption systems in regions where the
extragalactic background radiation dominates over the ionizing flux
from high--mass stars.  Unfortunately, the environment of the $z=1.04$
absorption system is unknown (see \S~\ref{sec:data}).  At $z\sim1$,
the extragalactic background radiation is a factor of $7$ times larger
than at $z\sim0$ \citep{haardtmadau96}.  However, the ionizing flux close to an
extreme starburst galaxy could dominate over the Haardt and Madau
spectrum.  Such a galaxy would have a photon flux of $10^{54}s^{-1}$,
of which $\sim1\%$ escape \citep{hjd97}.  The stellar flux would
strongly dominate Haardt and Madau (with a flux $10$ times larger at
$1$ Rydberg) only within $6$~kpc of the center of an extreme
starburst.  Since the absorption is relatively weak in the $z=1.04$
system, we consider this unlikely.  However, as an example, we explore
the effects of starburst models on the ionization parameter and
metallicity we infer for the {\MgII} clouds in Model 2 for subsystem A.

Instantaneous starburst models with ages of $0.01$ Gyr and $0.1$ Gyr
were combined with the extragalactic background.  Starburst model
spectra, which assume solar metallicity and a Salpeter IMF were taken
from \citet{bc93}.  We consider extreme models with a strength of $10$
times that of Haardt and Madau at $1$ Rydberg.

The $0.01$ Gyr model spectral shape is characterized by the sharp
edges of {\HI}, {\HeI} and {\HeII}.  These edges influence the balance
between the different ionization states of various metal line
transitions.  Specifically, we find more {\CII} and less {\SiIII} and
{\SiIV} than in the pure Haardt and Madau case. Therefore, the
inferred upper limit on the ionization parameter would be slightly
higher and a more extreme abundance pattern adjustment would be needed
for carbon.

Also, the ionizing spectrum in the $0.01$ Gyr model, has a relatively
large number of photons above the {\HI} edge.  Since these photons
have the largest cross section for ionizing hydrogen, the clouds will
be more ionized than in the pure Haardt Madau case.  Therefore, the
inferred metallicity is $\sim0.5$ dex lower.

A sharp Lyman edge characterizes the $0.1$ Gyr model.  However, unlike
the $0.01$ Gyr model, the spectrum is rather flat over higher
energies.  This flat spectrum is quite similar to the Haardt Madau
spectrum and therefore there is little change in the ionization
balance between the metal line transitions.  The inferred metallicity
for this model would be slightly higher ($\sim0.2$ dex) in this case.
This is due to a slightly softer ionizing spectrum for energies just
above the Lyman edge.

From this investigation of the effect of spectral shape we conclude
that our general results based upon a pure Haardt Madau spectrum are
likely to apply.

\section{Summary of Results}
\label{sec:summary}

In the previous section, we gave a detailed description of how
we constrained physical conditions for the multi-cloud, weak {\MgII}
absorber at $z=1.04$ along the line of sight to PG~$1634+706$.
Here, we summarize those results, present an overview of
a consistent model for the absorber, and compare to the results
of \citet{anticip}, which were based on lower resolution spectra:

\begin{enumerate}

\item{

Two subsystems, A and B, detected in low ionization transitions, are
separated by $\sim150$~{\kms}.  Assuming, for simplicity, that each
cloud in the subsystem has the same metallicity, then the observed
partial Lyman limit break is consistent with a metallicity of $\log
Z\sim-1.4$.  Models with higher metallicities and dust depletion
are not viable because carbon and silicon are depleted less
than magnesium, and they are already overproduced by models
with a solar abundance pattern.
Subsystem A has clouds with an ionization parameter of
$\log U\sim -2.3$, however, the {\MgII} could either be produced in
those clouds (Model 1) or in additional lower ionization clouds, $\log
U\sim -3.1$ (Model 2).  The ionization parameter of the clouds in
subsystem B are slightly lower, $\log U\sim -2.7$.}

\item{
The issue of whether subsystem A, centered at $v=0$~{\kms}, has such
an additional low ionization phase hinges partially on the question of
whether the {\MgII} and {\SiIV} profiles are kinematically aligned.
Figure~\ref{fig:offset} shows that, although
the shape of the {\SiIV}~$\lambda 1394$ absorption profile is similar
to the {\MgII} profile in subsystem A, it appears to be offset to the
red by a few kilometers per second. An
offset of the same magnitude was observed for one of the clouds in the
$z\sim1.9$ absorber along the line of sight to the HDF-South quasar,
where {\CIV}, {\SiIII}, and {\SiIV} are offset from {\CII} and {\SiII}
by $\sim4$~{\kms} \citep{DOdorico}. \citet{DOdorico} interpret this
offset as the consequence of an {\HII} region flow.  A similar offset
between {\SiII} versus {\CII} and {\SiIII} may exist in the
$z=0.00530$ absorber toward 3C~273 \citep{tripp3c273}.  The
interpretation of the $z\sim1.04$ system is more complex because
several clouds would be offset, but a similar explanation is
possible.}

\item{
The cloud sizes may also provide insight into whether a
one (Model 1) or two (Model 2) phase model of subsystem A is
more plausible.  If the {\MgII}
and {\SiIV} arise in the same phase, the four components of
subsystem A have sizes ranging from $6$--$24$~kpc.  It is
difficult to conceive of a physical picture in which four
such large structures would lie along the line of sight,
within $50$~{\kms}.  For a two phase model (with the offset
{\SiIV} components), the cloud sizes of the components responsible
for the {\SiIV} are reduced to several kpcs, with smaller
($100$--$300$~pc) {\MgII} clouds also contributing to the absorption.
This does seem more plausible in relationship to basic features
of galactic interstellar gas.}

\item{
Although both the offset {\SiIV} clouds and the clouds centered on
{\MgII} contribute to the {\SiIII} absorption profile, they do not
account for its smooth, broad shape.  This may indicate that much of
the {\SiIII} is produced by collisional ionization at a temperature
of $\log T\sim4.6$.}

\item{
Two broad {\OVI} absorption profiles are observed, one $50$~{\kms} to
the red of subsystem A and the other $50$~{\kms} to the blue of
subsystem B.  Assuming the {\OVI} and {\Lya} arise in the same phase,
the measured Doppler parameter of {\Lya} associated with the {\OVI}
places an upper limit on the temperature such that the {\OVI} could
not be produced by collisional ionization.  Both the {\OVI} and the
{\Lya}, which could not be fully produced by the subsystems, can arise
in the same photoionized phase.  The low resolution {\CIV} profile,
observed by FOS/HST, also requires a contribution from the blue broad
component.  These highly--ionized clouds ($-1.5 < \log U < -0.5$)
have metallicities larger than the low--ionization subsystems.
The model cloud sizes range from 10's to 100's of kpc, depending
on the metallicity, with larger metallicities leading to smaller
sizes.
}

\end{enumerate}

\subsection{Comparison with Previous Predictions}
\label{sec:compareprev}

\citet{anticip} studied the $z=1.04$ system before the STIS/HST
spectrum was released, using the low resolution FOS/HST spectra in conjunction
with the same HIRES/Keck spectra as were analyzed here.
Now, using the higher resolution STIS/HST UV data, we confirm
most of the basic results of that paper, and we are able to
resolve some ambiguities that were presented there.
The metallicity constraint is confirmed, along with the presence
of an additional high ionization cloud $\sim 200$~{\kms} to the blue
(corresponding to the blue broad component in the present models).  The metallicity
agreement is not surprising, since in both cases the constraint came from
HIRES/Keck spectra covering {\MgII} in conjunction with the
Lyman limit break whose appearance is not sensitive to resolution.
The presence of an additional high ionization component is
convincingly confirmed by the presence of {\OVI}
absorption at a consistent velocity.  It seems likely that
the {\CIV} seen in the FOS spectrum and the {\OVI} seen in
the STIS spectrum would arise in the same phase.

Because of the limitations of the low resolution data,
the \citet{anticip} work was unable to distinguish between
two scenarios for the origin of the {\CIV} in the redward part
of the profile.  Based on those data, it could have arisen either in the
same clouds with the {\MgII} or in a separate phase.  Now, because
the {\SiIV} profile has narrow components resolved
in the high resolution STIS data, we can see that the {\CIV} arises
either in the {\MgII} clouds (Model 1) or in the separate {\SiIV}
clouds (Model 2.)  It does not arise in the red broad component which is
responsible for producing {\OVI} and the red wing of {\Lya}.

The metallicities and ionization conditions of the blue broad component
were poorly constrained by \citet{anticip}, because only {\Lya}
and {\CIV} were available, and at low resolution.
They found that a model with $\log Z = -1.5$ and $\log U = -2.1$
could be consistent with the data, but higher metallicities were
also compatible, depending on the $b$ parameter assumed for the
unresolved broad component.  They indicated that low ionization
transitions might be observed at this same velocity.
Now, we find that the detection of strong {\OVI} indicates a higher
ionization parameter and a higher metallicity.  Also, although low
ionization transitions
are detected in the STIS spectrum, their absorption profiles are
offset by $\sim50$~{\kms} from the blue broad component.
The previous work therefore had the general idea that there
was absorption blueward of the main low ionization system, but
was able to discern little about the details.

\section{Discussion}
\label{sec:discussion}

From the derived physical constraints of the $z=1.04$ system along the
line of sight to PG~$1634+706$, we determined certain basic properties
that this absorber must have.  There are two separate subsystems, implying
that there are two distinct structures.  These subsystems
have lower metallicity, lower ionization regions, and kinematically
distinct, higher metallicity and higher ionization ``halos''.
Although these conclusions do not indicate a unique solution, they
suggest scenarios and types of structures that might be responsible.

{\it The subsystems of low-ionization material in the
$z=1.04$ system have a low metallicity ($\log Z\sim -1.4$)---}
Although giant galaxies would be expected to have lower metallicities at higher
redshifts than they do at present, a value as low as $\log Z\sim -1.4$ would be
more extreme than our expectations.  Almost all of the stars in the disk of the
Milky Way formed well before redshift one (coincident with the peak of the
cosmic star formation), and nearly all of them have metallicities greater
than $0.1$~solar, with values as large as solar being common \citep{wyse}.
It is consistent that the metallicities of the low ionization clouds
composing strong
{\MgII} absorbers at $z\sim1$ appear to have higher metallicities,
from 1/10th solar up to the solar value \citep{ding1634,ding1206}.

Low metallicities at $z\sim1$ are seen in damped {\Lya} absorbers
\citep{pettini}, and this is attributed to their hosts having a mix of
morphologies, commonly including dwarfs and low surface brightness
galaxies \citep{rao,bowendla}.  In Local Group dwarf galaxies at
present, metallicities range from $\log Z=-1.0$ to $-2.0$
\citep{mateo}.

The outer disks of spiral galaxies could also have low metallicities
consistent with our derived value for the $z=1.04$ system.  The iron
radial metallicity gradient in the Milky Way disk is $\sim
-0.06$~dex~kpc$^{-1}$, out to a radius of $\sim 16$~kpc
\citep{metgrad}.  We would expect weak {\MgII} absorption (as opposed
to strong) to arise only at galactocentric radii greater than $30$--$40$~kpc.
An extrapolation of the observed metallicity gradient would therefore
produce a low metallicity for radii at which weak {\MgII} absorption might
be expected.

{\it The broad, higher ionization components in the $z=1.04$ system have a higher
metallicity than the low ionization subsystems---}
The subsystem absorption is produced in higher density regions than the broad higher
ionization components.  Also, the individual subsystem clouds are narrower and
smaller (sizes of $\sim 1$~kpc) than the high ionization clouds (sizes of tens
to hundreds of kpcs).  This suggests that the subsystems arise in a smaller
region internal to a galaxy or structure, more likely to be close to the
stellar sources of metals.  If so, at face value, it is puzzling why the
more diffuse, highly ionized gas would have a higher metallicity than the
concentrated regions.  One possibility would be a scenario in which relatively
undiluted supernova ejecta are ejected from the central regions of a galaxy into
a diffuse medium.  This is seen to some extent in the ejecta of nearby galactic
winds, observed with Chandra, although the metallicities of the wind material
in these local systems are only somewhat larger than those of
the interstellar material
from which they originate \citep{martin1569,martinproc}.

Another clue may result from the abundance pattern of the blue broad component.
If we tune its ionization parameter and metallicity
so as to fit the {\CIV} and {\OVI} absorption (along with {\Lya}),
the {\NV} absorption is severely overproduced.  A reduction of the abundance
of nitrogen relative to carbon and oxygen would bring the model in accord
with the data.  Such a nitrogen deficiency (about an order of magnitude)
is characteristic of the dwarf galaxies in the Local Group \citep{mateo}.
This pattern is consistent with primary production of nitrogen in a low
metallicity environment \citep{henry}.

{\it Each of the two subsystems of the $z=1.04$ absorber has several
narrow low ionization components spread
over tens of {\kms}, which may be related to
a broad, high ionization component offset by $50$~{\kms}---}
The kinematic spread of a subsystem is reminiscent of that
we would expect for a line of sight through a spiral
disk with some increase due to the effects of interactions (eg.
contribution from warps, tidal debris, or fountain material).
Strong {\MgII} absorbers often show subsystems with
kinematic spreads of tens of {\kms}, though they have larger
equivalent width and they also often have ``satellite clouds'' at
larger velocities which contribute to a larger kinematic spread
for the overall system \citep{cv01}.  In the case of the strong
{\MgII} absorbers, the kinematics of the strongest subsystem have
been demonstrated to be consistent with the expectations of thick
disk kinematics \citep{kinmod}.
A kinematic spread of tens of {\kms} is also similar to the
rotation velocities of the larger Local Group dwarf irregulars and
to the velocity dispersions of Local Group dwarf spheroidals \citep{mateo}.

Significant offsets between the low and high ionization gas
are not uncommon in the nearby universe.  For example, offsets are
typically $\sim 30$~{\kms} between the {\OVI} absorption detected in
the Large Magellanic cloud and its low ionization absorption, which is
associated with its interstellar medium \citep{howk}.  The LMC {\OVI}
is consistent with an origin in coronal gas, being similar in
strength and velocity dispersion to that associated with the Milky Way
\citep{howk,savageo6}.
The observed kinematics for each of the $z=1.04$ subsystems could
be due to a low ionization contribution from clumps in the disk or
internal region of a dwarf, which would be
surrounded by a broad {\OVI} phase corona or halo.  The measured
{\OVI} Doppler parameter of $b\sim20$--$30$~{\kms} is consistent with
a dwarf galaxy halo dispersion
\citep{mateo}.  The offset of $\sim50$~{\kms} of the
broad {\OVI} component from the low--ionization gas could be due to
a kinematic offset between central/disk and halo components of the
dwarfs.

Another possibility for the kinematic difference between the
low and high ionization components in the $z=1.04$ absorber
is a layer structure of
the cone surrounding a breakthrough superbubble or wind.
In the case of a starbursting dwarf, hot, X--ray emitting gas is
ejected at high velocity from the central region of the galaxy (see,
eg. \citet{read, ptak, dahlem}).  The edges of the cone surrounding
the X-ray gas are cooler and are likely to have condensations that
would give rise to lower ionization absorption.  In fact, strong
{\NaI} absorption is detected in absorption against the majority
of nearby starbursts \citep{heckman00}.
We would expect ``transition layers'' in between the hot and cooler regions.
In these regions, moderate to higher ionization absorption could occur.  The
broad {\OVI} absorption observed in the $z=1.04$ system could trace
such a transition layer in the interior.  As the line of sight passes
through the outer layers, one might expect to see a group of clouds
producing lower-ionization absorption that would be moving at a speed
lower than that of the interior of the wind.  

{\it The low ionization absorption in the two $z=1.04$ absorber
subsystems is quite weak.  Each of them would be classified as a multiple
cloud, weak {\MgII} absorber---}
Because of the weak, low ionization absorption, statistically, we would infer
that neither of the two subsystems would be produced by a line of sight through
the inner $30$--$40$~kpc of a giant galaxy.  Unfortunately, in the case of
PG~$1634+706$ we have no direct information from imaging of the field
(as described in \S~\ref{sec:data}).

We can gain insight by considering the nature of the host galaxies or
structures for nearby weak {\MgII} absorbers, for which imaging can
reach down reasonably far on the dwarf galaxy luminosity function.
\citet{rosenberg} have proposed that the weak low ionization absorber
at $z=0.00530$ toward 3C~273 is produced in a thin shell around a
$M_B = -14.5$ ``post--starburst'' galaxy that has been found at an
impact parameter of $71 h_{70}^{-1}$~kpc.  This is consistent with
\citet{tripp3c273}'s speculations about that same system, and with
the theoretical ideas of \citet{theuns} concerning the formation of
condensations in dwarf galaxy winds.  \citet{rosenberg} have proposed
that this is likely to be a common origin of weak, low ionization
absorption systems.  There is another weak, low ionization
absorber found at $z\sim 0.005$ along the RXJ~$1230.8+0115$ sightline,
which is adjacent to the 3C~273 sightline, at a separation of
$\sim 350h_{70}^{-1}$~kpc \citep{rosenberg}.  However, in that case, the galaxy
responsible has not yet been identified.  There is a candidate host
dwarf galaxy known for the multiple--cloud, low ionization absorber at
$z=0.051$ toward PG~$1211+143$, but its star formation history
is unknown (2003, J. Stocke, private communication)

{\it The $z=1.04$ absorber has a possible offset between the {\MgII} and
{\SiIV} in subsystem A, and the {\SiIII} in that same subsystem
implies the presence of a collisionally ionized layer at intermediate temperature---}

Offsets between low to intermediate ionization transitions have been
seen in other absorption systems, as was noted in
\S~\ref{sec:summary}.  The interpretation could be a layered
structure, with different kinematics for the layers responsible for
the different ionization stages.  In the case of subsystem A, the
offsets in multiple components would require a series of such
structures with a common orientation.  The presence of a collisionally
ionized layer at an intermediate temperature ($T\sim100,000$~K) has
also been suggested for other $z\sim1$ absorption systems
\citep{ding1634,ding1206}.  This may also be related to shocks and
is consistent with a complex, layered structure.

{\it Two relatively strong, broad {\OVI} absorption features are present in the
$z=1.04$ absorber, each with a $50$~{\kms} offset from one of the subsystems---}

\citet{tripp1} have shown that most of the baryons in the universe are
trapped in a warm/hot phase traced by {\OVI}.  This has been
interpreted as intragroup gas, however, the kinematic spread of the
{\OVI} in some of the absorbers is not as large as would be
expected for a poor galaxy group \citep{tripp2,zabludoff}.  It is also
possible that such {\OVI} arises in the gaseous halo of a single galaxy
\citep{tripp2}. Some of the stronger {\OVI} systems have associated weak lower
ionization absorption \citep{savage02}.  It follows that there is some overlap
between the stronger {\OVI} absorbers seen by \citet{tripp1} and multiple
cloud weak {\MgII} absorbers such as the one studied here.

More generally, the kinematic spreads of the broad {\OVI} components
in the $z=1.04$ absorber are consistent with the velocity dispersions
of the larger Local Group dwarf galaxies \citep{mateo}.  They are also
consistent with the kinematics of the coronal gas around the Milky Way
and the LMC \citep{howk,savageo6}, as mentioned above.
The {\OVI} would also arise naturally in a wind model, somewhere between the
central X--ray emitting
gas and the cone in which the low ionization absorption would arise.

{\it The two subsystems of the $z=1.04$ absorber are separated by
$\sim 150$~{\kms}, and they appear to have fairly similar physical conditions.}

So far, we have discussed possible origins of the individual subsystems
in the $z=1.04$ absorber (and their possibly associated high ionization
broad components), but it is important to consider how two such
systems could arise.  The two subsystems are kinematically distinct,
i.e. they are well separated in velocity space. 
There are many scenarios that could give rise to absorption components
separated by $\sim 150$~{\kms}, but most can be categorized into one
of two themes.  In the first theme, there are two separate structures
along the line of sight that exist within the same group or halo.
In the second theme, a superbubble or wind is responsible for
producing two kinematic structures that arise in the same object.

To evaluate the idea that
two separate structures are responsible for the two subsystems, we
must consider the likelihood of intercepting two structures
along the same line of sight.  
The velocity separation of $150$~{\kms} between subsystems A and B is
characteristic of galaxy group velocity dispersions \citep{zabludoff}. 
The responsible structures might be outer disks
of galaxies or dwarf galaxies, as discussed above.  There are
many ways of considering this problem.  For example, if there are
typically ten satellite galaxies clustered within $200$~kpc
around a giant galaxy at $z\sim 1$, and if each satellite has
a cross section for weak low ionization absorption of $1200$~kpc$^2$,
there would be a $\sim 10~\%$ chance of a line of sight intercepting
one if it intercepted another.  In this scheme, we would expect
that most multiple--cloud weak {\MgII} absorbers would not have
two separate subsystems, i.e. we would expect that typically
only one satellite would be intercepted.  This seems consistent
with the data since, in a sample of $\sim 10$ multiple cloud,
weak {\MgII} absorbers, only in this one case are there two separate
subsystems \citep{weak1}.

If a superwind or bubble is responsible for the $z=1.04$ absorber,
we would expect that each of the two low-ionization
subsystems and its associated {\OVI} phase would represent an opposite side
of a wind.  The two sides could plausibly be separated by $v\sim150$~{\kms}
\citep{heckman00}.  There would have to be cold condensations with a
fairly large covering factor on both sides of the wind in order to see
two low ionization subsystems.  One could easily imagine other cases
with asymmetric blowouts or stochastic structure such that only one
side was detected or such that one side was only detected in high
ionization transitions.

The very strongest {\MgII} absorbers (which are common only at $z>1$),
could also have an origin in superwinds \citep{bond}.
These absorbers have a
characteristic saturated ``double--trough'' absorption profile in {\MgII},
which breaks up into multiple clouds in other weaker low-ionization
transitions.  As with our two subsystems, the saturated troughs are
typically separated by $\sim100$--$200$~{\kms}.
Strong low-ionization absorption could arise from clouds close to the
point of origin of the wind or from early evolutionary stages of the
superwind phenomenon.  As we reach larger distances or later stages
when low--ionization superwind clouds may have fragmented or
dispersed, we might expect weaker absorption profiles like the ones we
are seeing in the $z=1.04$ absorber toward PG~$1634+706$.

If more multiple cloud, weak {\MgII} absorbers with two subsystems are observed
it should be possible to test between an origin in two structures and an origin
in a wind or outflow.  If winds are responsible then we
would expect to see broad, high ionization absorption to the blue of the blueward subsystem
and to the red of the redward subsystem.
If two separate structures were responsible, the offset low and high
ionization components would be due to kinematic differences between two components
in such structures.  In such a case, the broad, high ionization absorption should
arise at random to the red or to the blue of each of the subsystems, and sometimes
would even be superimposed.

{\it There are about two thirds as many multiple cloud weak {\MgII}
absorbers as there are strong {\MgII} absorbers---}
There must be a set of structures responsible for multiple cloud
weak {\MgII} absorption with a cross section comparable to the
set of $\sim30$~kpc regions around all $L^*$ galaxies.
If these are outer disks around all $>0.05L^*$ galaxies, an annulus with
thickness of $\sim 10$~kpc would be required.

\citet{bond} found that it was plausible for the expected number of
superwinds at redshift $z\sim1.5$ to account for the observed number
of saturated ``double -- trough'' strong {\MgII} profiles.  Similarly,
we speculate whether the outer winds of a plausible number of bursting
dwarfs could account for the observed number of multiple-cloud weak
{\MgII} absorbers.

There are about 2/3 as many multiple cloud weak {\MgII} absorbers are
there are strong {\MgII} absorbers \citep{weak1,weak2}.  Strong
{\MgII} absorption arises from regions around $>0.05 L^*$ galaxies with
sizes of $\sim30h^{-1}$~kpc
\citep{bb91,bergeron92,lebrun93,sdp94,steidel95,3c336}.  If the regions
around starbursting dwarfs that give rise to multi-cloud weak {\MgII}
absorption have sizes of $10h^{-1}$~kpc, then a number density $6$
times that of $\sim L^*$ galaxies could account for all these
multi-cloud weak {\MgII} absorbers.

Realistically, it seems likely that there would be contributions to multiple cloud
weak {\MgII} absorption from a variety of types of structures and processes.
Conversely, all types of structures that contain gas must all make some
contribution to the absorber population, contributing either strong or weak
{\MgII} absorption and/or contributing to {\CIV} or {\OVI} absorption.
It is fairly well understood that $L^*$ galaxies produce most all
of the observed strong {\MgII} absorption at $z\sim1$.  But it is not
well understood what type of absorption systems quiescent dwarf galaxies or
dwarf galaxies with winds would produce.

Above, we have considered the various properties of the $z=1.04$ absorber
and discussed what types of structures and processes might be consistent.
Clearly, there is no unique interpretation.  However, a model
in which this absorber is related to dwarf galaxy winds is consistent
with most of the listed properties.  It can explain the phase structure
of the absorber, the relative metallicities of the phases,
and the kinematics of the two subsystems relative to the broad, higher
ionization components.  This model is also appealing in that it
accounts for some of the absorption cross section presented by
starbursting dwarfs, which were common at redshift one.

Many more multiple cloud, weak {\MgII} absorbers must be studied in
order to evaluate the relative contributions of dwarf galaxy winds, of
quiescent dwarf galaxies, of outer galaxy disks, and of other phenomena.  In conjunction with
such a statistical study, it will be essential to find analogs
in the local universe so that the host galaxies and responsible
processes can be directly identified.


\acknowledgements  
Support for this work was provided by the NSF (AST-9617185) and by NASA
(NAG5-6399 and STSI GO-08672.01-A).  SGZ was supported by an NSF REU supplement. 



\newpage

\newpage

\clearpage

\clearpage

\clearpage

\clearpage

\clearpage

\clearpage

\clearpage

\begin{deluxetable}{lrrrrrrrrrrrrr}
\label{tab:tab1}
\tablenum{1}
\tabletypesize{\footnotesize}
\tablewidth{0pt}
\tablecaption{Equivalent Widths}
\tablehead{
\colhead{Transition}   &
\colhead{Subsystem A}   &
\colhead{Subsystem B}   &
\colhead{Total}

}

\startdata
{\Lya}~$\lambda 1215$ & $---$ & $---$ & $1.45 \pm 0.01$\\
{\Lyb}~$\lambda 1026$ & $---$ & $---$ & $1.09 \pm 0.03$\\
{\Lyg}~$\lambda 972$ & $---$ & $---$ & $0.891 \pm 0.04$\\
{\Lyd}~$\lambda 950$ & $---$ & $---$ & $0.859 \pm 0.03$\\
{\Lye}~$\lambda 923$ & $---$ & $---$ & $0.611 \pm 0.04$\\
{\MgII}~$\lambda 2796$ & $0.097 \pm 0.008$ & $---$ & $0.097 \pm 0.008$\\
{\MgII}~$\lambda 2803$ & $0.046 \pm 0.009$ & $<0.006$ & $<0.052$\\
{\FeII}~$\lambda 2600$ & $<0.038$ & $<0.038$ & $<0.076$\\
{\SiII}~$\lambda 1260$ & $0.024 \pm 0.005$ & $<0.008$ & $<0.032$\\
{\CII}~$\lambda 1335$ & $0.067 \pm 0.007$ & $0.017 \pm 0.004$ & $0.084 \pm 0.008$\\
{\SiIII}~$\lambda 1207$ & $0.28 \pm 0.007$ & $0.14 \pm 0.006$ & $0.42 \pm 0.009$\\
{\SiIV}~$\lambda 1394$ & $0.20 \pm 0.006$ & $0.055 \pm 0.004$ & $0.255 \pm 0.007$\\
{\SiIV}~$\lambda 1403$ & $0.060 \pm 0.006$ & $0.024 \pm 0.005$ & $0.084 \pm 0.008$\\
{\NV}~$\lambda 1239$ & $<0.010$ & $<0.010$ & $<0.020$\\
{\OVI}~$\lambda 1032$ & $0.094 \pm 0.01$ & $0.14 \pm 0.01$ & $0.23 \pm 0.01$\\
{\OVI}~$\lambda 1038$ & $0.14 \pm 0.01$ & $0.092 \pm 0.01$ & $0.23 \pm 0.01$\\
{\CIV}~$\lambda 1548$ & $---$ & $---$ & $0.40 \pm 0.02$\\
{\CIV}~$\lambda 1551$ & $---$ & $---$ & $0.38 \pm 0.02$\\

   \hline
\enddata
\vglue -0.05in
\tablecomments{
\baselineskip=0.7\baselineskip
The equivalent widths were measured for various transitions detected
at the $3$ sigma level. For transitions that were not detected, an $3$
sigma upper limit is given.  Because both subsystems are blended
together in the low resolution FOS data covering {\CIV}, only a total
equivalent width is shown.  Although covered in the higher resolution
STIS data, the two subsystems as traced by the {\Lya} are also blended
and individual components can not be measured.}

\end{deluxetable}
\clearpage

\newpage
\thispagestyle{empty}
\begin{deluxetable}{lrrrrrrrrrrrrr}
\label{tab:tab2}
\tablenum{2}
\tabletypesize{\footnotesize}
\rotate
\tablewidth{0pt}
\tablecaption{Photoionized Cloud Properties}

\tablehead{
\colhead{} &
\colhead{$v$}  &
\colhead{$N_{\rm tot}$}   &
\colhead{$N({\HI})$} &
\colhead{$\log U$} &
\colhead{$Z$} &
\colhead{size} &
\colhead{$N({\MgII})$} &
\colhead{$N({\SiII})$} &
\colhead{$N({\CII})$} &
\colhead{$N({\SiIII})$} &
\colhead{$N({\SiIV})$} &
\colhead{$N({\CIV})$} &
\colhead{$b$} \\
\colhead{} &
\colhead{[{\kms}]} &
\colhead{[{\cmsq}]} &
\colhead{[{\cmsq}]} &
\colhead{} &
\colhead{[$Z_{\odot}$]} &
\colhead{[kpc]} &
\colhead{[{\cmsq}]} &
\colhead{[{\cmsq}]} &
\colhead{[{\cmsq}]} &
\colhead{[{\cmsq}]} &
\colhead{[{\cmsq}]} &
\colhead{[{\cmsq}]} &
\colhead{[{\kms}]}
}

\startdata

\multicolumn{14}{c}{\sc Subsystem A - Model 1}\\
\hline
{\MgII}$_{\rm 1}$ & $-35.4$ & $19.5$ & $16.4$ & $-2.3$ & $-1.4$ & $8$ & $11.45$ & $11.5$ & $12.6$ & $12.6$ & $12.3$ & $13.4$ & $5.82$ \\
{\MgII}$_{\rm 2}$ & $-12.7$ & $19.9$ & $16.6$ & $-2.5$ & $-1.4$ & $6$ & $11.90$ & $11.8$ & $12.9$ & $12.9$ & $12.4$ & $13.3$ & $10.78$ \\
{\MgII}$_{\rm 3}$ &   $2.6$ & $20.0$  & $16.8$ & $-2.3$ & $-1.4$ & $24$ & $12.02$ & $12.0$ & $13.1$ & $13.1$ & $12.7$ & $13.8$ & $3.07$ \\
{\MgII}$_{\rm 4}$ &  $18.3$ & $19.7$ & $16.5$ & $-2.3$ & $-1.4$ & $13$ & $11.68$ & $11.7$ & $12.8$ & $12.8$ & $12.5$ & $13.6$ & $5.92$ \\

\hline
\multicolumn{14}{c}{\sc Subsystem A - Model 2}\\
\hline
{\MgII}$_{\rm 1}$ & $-35.4$ & $18.3$ & $16.1$ & $-3.1$ & $-1.4$ & $0.09$ & $11.45$ & $11.1$ & $12.2$ & $11.9$ & $10.9$ & $11.4$ & $5.82$ \\
{\MgII}$_{\rm 2}$ & $-12.7$ & $18.8$ & $16.5$ & $-3.1$ & $-1.4$ & $0.2$ & $11.90$ & $11.5$ & $12.6$ & $12.3$ & $11.4$ & $11.8$ & $10.78$ \\
{\MgII}$_{\rm 3}$ &   $2.6$ & $18.9$  & $16.6$ & $-3.1$ & $-1.4$ & $0.3$ & $12.02$ & $11.6$ & $12.8$ & $12.4$ & $11.5$ & $11.9$ & $3.07$ \\
{\MgII}$_{\rm 4}$ &  $18.3$ & $18.6$ & $16.3$ & $-3.1$ & $-1.4$ & $0.1$ & $11.68$ & $11.3$ & $12.9$ & $12.1$ & $11.2$ & $12.1$ & $5.92$ \\

{\SiIV}$_{\rm 1}$ & $-30.3$ & $19.0$ & $15.8$ & $-2.3$ & $-1.4$ & $2$ & $10.9$ & $11.4$ & $12.1$ & $12.5$ & $12.2$ & $12.9$ & $5.7$ \\
{\SiIV}$_{\rm 2}$ & $-7.8$  & $18.9$ & $15.8$ & $-2.3$ & $-1.4$ & $2$ & $10.8$ & $11.4$ & $12.0$ & $12.5$ & $12.2$ & $12.8$ & $11.0$ \\
{\SiIV}$_{\rm 3}$ & $3.9$   & $19.3$ & $16.1$ & $-2.3$ & $-1.4$ & $5$ & $11.2$ & $11.7$ & $12.4$ & $12.8$ & $12.5$ & $13.2$ & $5.0$ \\
{\SiIV}$_{\rm 4}$ & $23.3$  & $19.1$ & $16.0$ & $-2.3$ & $-1.4$ & $3$ & $11.1$ & $11.6$ & $12.3$ & $12.7$ & $12.4$ & $13.1$ & $3.0$ \\

\hline
\multicolumn{14}{c}{\sc Subsystem B} \\
\hline

{\SiIII}$_{\rm 1}$ & $-184.5$ & $17.9$ & $15.3$ & $-2.7$ & $-1.4$ & $0.08$ & $10.5$ & $10.8$ & $11.9$ & $11.8$ & $11.2$ & $12.0$ & $12.75$ \\
{\SiIII}$_{\rm 2}$ & $-144.9$ & $18.9$ & $16.2$ & $-2.7$ & $-1.4$ & $0.8$ & $11.5$ & $11.8$ & $12.9$ & $12.8$ & $12.2$ & $13.0$ & $22.00$ \\
{\SiIII}$_{\rm 3}$ & $-127.7$ & $18.5$ & $15.8$ & $-2.7$ & $-1.4$ & $0.3$ & $11.1$ & $11.4$ & $12.5$ & $12.4$ & $11.8$ & $12.6$ & $6.48$ \\

\hline
\enddata
\vglue -0.05in
\tablecomments{
\baselineskip=0.7\baselineskip
Cloud properties are summarized for example of Model 1 (just 4 {\MgII}
clouds) and Model 2 (4 {\MgII} clouds and 4 offset {\SiIV} clouds) for
subsystem A.  Subsystem B properties are also given.  For both models,
the abundances of silicon and carbon for the {\MgII} clouds of
subsystem A are adjusted downwards from solar by $0.5$~dex.  All
column densities are indicated as logs of the values in units of
{\cmsq}.  Column densities of the {\MgII} and {\SiIV} clouds from
subsystem A as well as the {\SiIII} clouds from subsystem B were
determined by Voigt-profile fits.  Also included is the total Hydrogen
column density, $N_(tot)$, (including {\HI} and {\HII}).  The velocities,
also determined by VP fits, are measured relative to the median of the
apparent optical depth distribution for the {\MgII} clouds.  Doppler
parameters listed in the last column correspond to the transition for
which that cloud is optimized (listed in the first column).  The
models summarized in this table are also shown in
Figures~\ref{fig:model1a}, \ref{fig:model1b}, \ref{fig:model2a} and
\ref{fig:model2b}.  }

\end{deluxetable}
\clearpage

\newpage
\thispagestyle{empty}
\begin{deluxetable}{lrrrrrrrrrrrrr}
\tablenum{3}
\label{tab:tab2}
\tabletypesize{\footnotesize}
\rotate
\tablewidth{0pt}
\tablecaption{Photoionized Cloud Properties}
\tablehead{
\colhead{} &
\colhead{$v$}  &
\colhead{$N_{\rm tot}$}   &
\colhead{$N({\HI})$} &
\colhead{$\log U$} &
\colhead{$Z$} &
\colhead{size} &
\colhead{$N({\CIV})$} &
\colhead{$N({\NV})$} &
\colhead{$N({\OVI})$} &
\colhead{$b$} \\
\colhead{} &
\colhead{[{\kms}]} &
\colhead{[{\cmsq}]} &
\colhead{[{\cmsq}]} &
\colhead{} &
\colhead{[$Z_{\odot}$]} &
\colhead{[kpc]} &
\colhead{[{\cmsq}]} &
\colhead{[{\cmsq}]} &
\colhead{[{\cmsq}]} &
\colhead{[{\kms}]}
}

\startdata
\multicolumn{11}{c}{\sc Broad Components}\\
\hline
                
{\HI} & $-210.2$ & $18.9$ & $15.0$ & $-1.5$ & $-0.5$ & $12$ & $14.3$ & $13.0$ & $14.4$ & $27$ \\
{\HI} & $40.3$ & $19.2$ & $14.0$ & $-0.5$ & $-1.0$ & $257$ & $12.5$ & $12.8$ & $14.2$ & $37$ \\

\hline
\enddata
\vglue -0.05in

\tablecomments{
\baselineskip=0.7\baselineskip
All column densities are indicated as logs of the values in units of
{\cmsq}.  Column densities of the two broad components were determined
by Voigt-profile fits to {\Lya}.  Doppler parameters listed in the
last column correspond to the transition for which that cloud is
optimized (listed in the first column).  The values shown in this
table represent a plausible model for the broad components, using
Model 2 for subsystem A, also shown in Figure~\ref{fig:broad}.  The
blue broad component has an abundance pattern in which nitrogen is
$0.8$~dex deficient relative to other elements.  The choice of model
for subsystem A only slightly affects the parameters for the blue
broad component. }

\end{deluxetable}
\clearpage


\begin{thebibliography}{XXX}

\bibitem[Bahcall {\etal}(1993)]{cat1}
Bahcall, J. N., {\etal} 1993, ApJS, 87, 1 

\bibitem[Bahcall {\etal}(1996)]{cat2}
Bahcall, J. N., {\etal} 1996, ApJ, 457, 19

\bibitem[Bergeron \& Boiss\'{e}(1991)]{bb91}
Bergeron, J., \& Boiss\'{e}, P. 1991, A\&A, 243, 344

\bibitem[Bergeron {\etal}(1992)]{bergeron92}
Bergeron, J., Cristiani, S., \& Shaver, P. A. 1992, A\&A, 257, 417

\bibitem[Bond {\etal}(2001)]{bond}
Bond, N. A., Churchill, C. W., Charlton C. C., \& Vogt, S. S., 2001,
ApJ, 562, 641

\bibitem[Bowen, Tripp, \& Jenkins(2001)]{bowendla}
Bowen, D. V., Tripp, T. M., \& Jenkins, E. B. 2001, AJ, 121, 1456

\bibitem[Brown {\etal}(2002)]{stis1}
Brown, T. {\etal} 2002, HST STIS Data Handbook, version 4.0, ed.. Mobasher, Baltimore, STScI

\bibitem[Bruzual \& Charlot(1993)]{bc93}
Bruzual, A. G., \& Charlot, S. 1993, ApJ, 405, 538

\bibitem[Charlton \& Churchill(1998)]{kinmod}
Charlton, J. C., \& Churchill, C. W. 1998, ApJ, 499, 181

\bibitem[Charlton {\etal}(2000)]{anticip}
Charlton, J. C., Mellon, R. R., Rigby, J. R., \& Churchill, C. W. 2000 ApJ, 545, 645

\bibitem[Charlton {\etal}(2003)]{weak1634}
Charlton, J. C., Ding, J., Zonak, S. G., Churchill, C. W., Bond, N. A., \&
Rigby, J. R. 2003, ApJ, 589, 311

\bibitem[Chen, Hou, \& Wang(2003)]{metgrad}
Chen, L., Hou, J. L., \& Wang, J. J. 2003, AJ, 125, 1397



\bibitem[Churchill {\etal}(2000)]{archiveI}
Churchill, C. W., Mellon, R. R., Charlton, J. C., Jannuzi, B. T.,
Kirhakos, S., Steidel, C. C., \& Schneider, D. 2000, ApJS, 130, 91

\bibitem[Churchill {\etal}(1999)]{weak1}
Churchill, C. W., Rigby, J. R., Charlton, J. C., \& Vogt, S. S. 1999,
ApJS, 120, 51

\bibitem[Churchill \& Vogt(2001)]{cv01}
Churchill, C. W., \& Vogt, S. S. 2001, AJ, 122, 679



\bibitem[Dahlem, Weaver, \& Heckman(1998)]{dahlem}
Dahlem, M., Weaver, K. A., \& Heckman, T. M. 1998, ApJS, 118, 401

\bibitem[D'Odorico \& Petitjean(2001)]{DOdorico}
D'Odorico, V., \& Petitjean, P. 2001, A\&A, 370, 729

\bibitem[Ding {\etal}(2003a)]{ding1634}
Ding, J., Charlton, J. C., Zonak, S. G., \& Churchill, C. W. 2003a,
ApJ, 587, 551

\bibitem[Ding {\etal}(2003b)]{ding1206}
Ding, J., Charlton, J. C., Churchill, C. W., \& Palma, C. 2003b,
ApJ, 590, 746

\bibitem[Farrah {\etal}(2002)]{farrah}
Farrah, D., Verma, A., Oliver, S., Rowan--Robinson, M. \& McMahon, R. 2002,
MNRAS, 329, 605

\bibitem[Ferland(2001)]{ferland}
Ferland,~G. 2001, Hazy, A Brief Introduction to Cloudy 96.00

\bibitem[Haardt \& Madau(1996)]{haardtmadau96}
Haardt, F., \& Madau, P. 1996, ApJ, 461, 20

\bibitem[Heckman {\etal}(2000)]{heckman00}
Heckman, T. M., Lehnert, M. D., Strickland, D. K., \& Armus, L. 2000,
ApJS, 129, 493


\bibitem[Henry, Edmunds, \& K\"oppen(2000)]{henry}
Henry, R. B. C., Edmunds, M. G., \& K\"oppen, J. 2000, ApJ, 541, 660

\bibitem[Howk {\etal}(2002)]{howk}
Howk, J. C., Sembach, K. R., Savage, B. D., Massa, D., Friedman, S. D.,
\& Fullerton, A. W. 2002, ApJ, 569, 214

\bibitem[Hurwitz, Jelinsky, \& Dixon(1997)]{hjd97}
Hurwitz, M., Jelinsky, P., Dixon, W. V. D. 1997, ApJ, 481 L31

\bibitem[Jannuzi {\etal}(1998)]{cat3}
Jannuzi, B. T., {\etal} 1998, ApJS, 118, 1

\bibitem[Kimble {\etal}(1998)]{kimble}
Kimble, R. A., {\etal} 1998, ApJ, 492, L83

\bibitem[Le~Brun {\etal}(1993)]{lebrun93}
Le~Brun, V., Bergeron, J., Boiss\'e, P., \& Christian, C. 1993,
A\&A, 279, 33

\bibitem[Martin(2003)]{martinproc}
Martin, C. L. 2003, in ``The IGM/Galaxy Connection: The Distribution of Baryons at $z=0$'',
eds. J. L. Rosenburg and M. E. Putman (Kluwer:Dordrecht), p. 205

\bibitem[Martin, Kobulnicky, \& Heckman(2002)]{martin1569}
Martin, C. L., Kobulnicky, H. A., Heckman, T. M., 2002, ApJ, 574, 663

\bibitem[Mateo(1998)]{mateo}
Mateo, M. 1998, ARAA, 36, 435



\bibitem[Pettini(2003)]{pettini}
Pettini, M. 2003, in ``Cosmochemistry: The Melting Pot of Elements'',
(Cambridge University Press: Cambridge), in press

\bibitem[Ptak {\etal}(1997)]{ptak}
Ptak, A., Serlemitsos, P., Yaqoob, T., Mushotzky, R.,\& Tsuru, T.
1997, AJ, 113, 1286

\bibitem[Rao \& Turnshek(2000)]{rao}
Rao, S. M., \& Turnshek, D. A. 2000, ApJS, 130, 1


\bibitem[Read, Ponman, \& Strikland(1997)]{read} 
Read, A. M., Ponman, T. J., \& Strickland, D. K. 1997, MNRAS, 286, 626

\bibitem[Rigby {\etal}(2002)]{weak2}
Rigby, J. R., Charlton, J. C. \& Churchill, C. W., 2002, ApJ, 565, 743

\bibitem[Rosenberg {\etal}(2003)]{rosenberg}
Rosenberg, J. L., Ganguly, R., Giroux, M. L., \& Stocke, J. T. 2003,
ApJ, in press


\bibitem[Savage {\etal}(2002)]{savage02}
Savage, B. D., Sembach, K. R., Tripp, T. M., \& Richter, P. 2002,
ApJ, 564, 631

\bibitem[Savage {\etal}(2003)]{savageo6}
Savage, B. D., Sembach, K. R., Wakker, B. P., Richter, P., Meade, M.,
Jenkins, E. B., Shull, J. M., Moos, H. W., \& Sonneborn, G. 2003,
ApJS 146, 125

\bibitem[Schneider {\etal}(1993)]{dpsKP}
Schneider {\etal} 1993, ApJS, 87, 45

\bibitem[Steidel(1995)]{steidel95}
Steidel, C.C. 1995, in QSO Absorption Lines, ed.\ G. Meylan 
(Garching : Springer Verlag), 139

\bibitem[Steidel {\etal}(1997)]{3c336}
Steidel. C. C., Dickinson, M., Meyer, D. M., Adelberger, K. L., \&
Sembach, K. R. 1997, ApJ, 480, 568

\bibitem[Steidel, Dickinson, \& Persson(1994)]{sdp94}
Steidel, C. C., Dickinson, M. \& Persson, E. 1994, ApJ, 437, L75 

\bibitem[Steidel {\etal}(2002)]{steidel02}
Steidel, C. C., Kollmeier, J. A., Shapley, A. E., Churchill, C. W.,
Dickinson, M., \& Pettini, M. 2002, ApJ, 570, 526

\bibitem[Steidel \& Sargent(1992)]{ss92}
Steidel, C. C., \& Sargent, W. L. W. 1992, ApJS, 80, 1

\bibitem[Sutherland \& Dopita(1993)]{sd93}
Sutherland, R.S., \& Dopita, M.A., 1993, ApJS, 88, 253

\bibitem[Theuns, Mo, \& Schaye(2001)]{theuns}
Theuns, T., Mo, H. J., \& Schaye, J. 2001, MNRAS, 321, 450

\bibitem[Tripp {\etal}(2002)]{tripp3c273}
Tripp, T. M., et al. 2002, ApJ, 575, 697

\bibitem[Tripp, Savage, \& Jenkins(2000)]{tripp1}
Tripp, T. M., Savage, B. D., \& Jenkins, E. B. 2000, ApJ, 534, L1

\bibitem[Tripp \& Savage(2000)]{tripp2}
Tripp, T. M., \& Savage, B. D. 2000, ApJ, 542, 42

\bibitem[Vogt {\etal}(1994)]{vogt94}
Vogt, S. S. et al. 1994, Proc. SPIE, 2198, 362

\bibitem[Vogt {\etal}(1995)]{vogt95}
Vogt, S. S., Mateo, M. Olszewski, E. W., Keane, M. J. 1995, AJ, 109, 151


\bibitem[Wyse \& Gilmore(1995)]{wyse}
Wyse, R. F. G., \& Gilmore, G. 1995, AJ, 110, 2771

\bibitem[Zabludoff \& Mulchaey(1998)]{zabludoff}
Zabludoff, A. I., \& Mulchaey, J. S. 1998, 498, 39

\end{thebibliography}
\end{document}